\documentclass[a4paper,11pt]{article}

\usepackage{jheppub_modified} 
\usepackage{graphicx}   
\usepackage[T1]{fontenc} 
\usepackage{slashed}
\usepackage{feynmp}
\usepackage{epsfig}
\usepackage{tikz}
\usepackage{enumerate}

\newcounter{exno}

\newcommand{\pt}{\partial}
\newcommand{\La}{\Lambda}
\newcommand{\om}{\omega}
\title{Lectures on Effective Field Theory}
\subheader{2015 NEXT Lectures}

\author{Ben Gripaios}
\affiliation{Cavendish Laboratory,\\JJ Thomson Avenue,
\\ Cambridge, 
CB3 0HE,
United Kingdom.\\ \\ \today}

\emailAdd{gripaios@hep.phy.cam.ac.uk}

\begin{document} 
\maketitle
\flushbottom

\acknowledgments
I thank my collaborators and the many lecturers who
have provided inspiration over the years; some of their excellent
notes can be found in the references. I thank M.~Redi and J.~Serra for
collaboration on the form of the gauge contributons to the composite Higgs potential. I also thank D. Sutherland for
many helpful comments and suggestions on how to improve the notes. The references are intended to
provide an entry point into the literature, and so are necessarily
incomplete; I apologize to those left out.
For errors and comments, please contact
me by e-mail at the address on the front page.

\section{Avant propos}
Let us suppose that we wish to describe some physical system on large
distance and time scales. Suppose, furthermore, that the system exhibits some kind of
random, local (or short-distance) fluctuations (for example, these fluctuations may be
the ones inherent in quantum mechanics). The formalism for describing such a system is called `effective
field theory' and is the subject of these lectures.

Note that it is already something of a miracle that such a theory
exists at all. Experience tells us that systems can be extremely
complicated on short-distance scales. Even though we are not so
arrogant as to try to
describe that short-distance physics, we know that that physics is
there and that it is what gives rise to the long-distance physics that
we {\em do} wish to describe. 

To give an example, consider QCD. Not quantum chromodynamics, but
quantum {\em cow} dynamics. Scientists now know that a cow, viewed at short-distance scales, is a very complicated object indeed, with multiple
stomachs made of cells made of proteins made of atoms made of
electrons and nuclei made of quarks made of goodness-knows-what. 
These quarks and electrons interact with each other (and with the
quarks and electrons in other cows) via the complicated quantum dynamics of
QED and QCD (the other, chromo, version). 

Viewed in this way, the problem of the computation of cow-cow
scattering looks like a very hard problem indeed. 

But viewed from far enough away (at large enough distance scales), a
cow behaves, for all intents and purposes,
like a point particle of mass $M$, with no internal dynamics at
all. Moreover, when we scatter 2 cows off each other, we see a very
simple, contact interaction (albeit with some rather complicated final
states, corresponding to inelastic scattering).

This example makes it clear that the desired miracle sometimes does
happen -- one doesn't need to know about gauge theory in order to
study long-distance cow-cow scattering. This is just as well, if you are a physicist. Indeed, I call
the miracle the `miracle of physics', because it is the basic reason
why physicists have ever been able to
make {\em any} progress and why physics enjoys the hegemony that it
does today: without the miracle, we could never get started on
tackling a physical system with a given length scale (e.g. on a desk
in a lab), without first worrying about all the other
physics taking place on all other distance scales throughout the Universe.

Enough philosophy. What are the ingredients of an effective field theory? Clearly, we need some
degrees of freedom. These will be represented by space-time
fields. The dynamics of the physical system may well be invariant
under some group of symmetries (such as space-time translations,
rotations, or Lorentz boosts), in which case we will need to specify
how the group acts on the
fields. We will then write the most general dynamics (in the form of
an action) for the fields that is invariant under the group action. 
We do this not because of a desire to be as general as possible;
rather, we will find that the short-distance fluctuations will, of its
own accord, generate the most general dynamics consistent with the
symmetries.\footnote{This is sometimes called `Gell-Mann's
  Totalitarian Principle': everything which is not forbidden is
  compulsory.}

You might be thinking that this sounds a lot like quantum field
theory (QFT). It is. In QFT, the lore is that one decides on the
fields and symmetries, and then writes down the most general {\em
  renormalizable} action for the fields that is consistent with the
symmetries. The insistence on renormalizability guarantees that one has
a theory which can be used to make predictions on {\em all} length
scales, including arbitrarily short ones. This is not only rather
arrogant, but also rather pointless, 
because no one has yet done an experiment on an arbitrarily short
distance scale! So EFT is really just the correct way to do
QFT. Unfortunately, it receives rather scant treatment in the QFT
textbooks. Fortunately, there are lots of excellent lecture notes
available \cite{Polchinski:1992ed,Manohar:1996cq,Rothstein:2003mp,Kaplan:2005es} and I encourage you to read as many of them as
possible. My goal here is not to repeat what others have said already,
but rather to give you the basic outline and then illustrate the
principles and pitfalls via several examples, namely the Standard
Model of particle physics, the non-linear sigma model, and the quantum
theory of perfect fluids. Other instructive examples that are
discussed in lecture notes elsewhere are the Euler-Heisenberg
lagrangian of low-energy QED, Landau's theory of Fermi liquids
\cite{Polchinski:1992ed}, and the effective theory of heavy quarks \cite{Manohar:1996cq,Rothstein:2003mp}.

\section{Notation and conventions}
As usual, $\hbar = c= 1$, and our metric is mostly\footnote{When we
  study the EFT of a perfect fluid in the last lecture, we'll switch
  to mostly-plus. Sorry!} mostly-minus: $\eta^{\mu
  \nu} = \mathrm{diag} (1,-1,-1,-1)$. We will exclusively use 2-component
left-handed Weyl
fermions.
In the Standard Model for example, the fermions are
$\psi \in \{q,u^c,d^c,l,e^c\}$. Kinetic terms are written as $ i
\overline{\psi} \overline{\sigma}^\mu \partial_\mu \psi$ and a Dirac mass
term for $\psi$ and $\chi$ is written as $\psi \cdot \chi +
\mathrm{h.\ c.}$. See \cite{Gripaios:2015gxa} for more details.

\section{Modus Operandi}
\subsection{QFT redux}
I assume that you know all about bog-standard QFT.\footnote{Only
  joking: no one knows {\em all} about QFT. But I hope that you at
  least know the basics.} There, the rules of the game are that we
decide upon a set of fields and a group of symmetries acting upon
them, and then write the most general renormalizable action involving
them. You well know, I hope, that in terms of their canonical
or engineering dimensions, this necessarily restricts us to terms in
the action of dimension four or less. The number of such terms is
finite (if the number of fields is). We assign each term an arbitrary
coefficient (though the coefficients of kinetic terms can be
set, without loss of generality, to
one, if the fields are complex, or one-half, if they are real).
Given that there is a finite number of such parameters ($n$ say), we have the
possibility of constructing a physical theory, in the sense that once
we have made $n$ suitable measurements to fix the values of the
parameters, we can start to make predictions for the results of other
measurements.

Things are not quite so straightforward in practice,
because when we try to fix the values of the {\em bare}  parameters, we find that
they have to be infinite. But in a renormalizable theory, these
infinites can be absorbed into finite, scale-dependent,  {\em
  renormalized} parameters, such that all relations between physical
observables
are finite, and we have a {\em bona fide} physical theory. 

The appearance of infinities nevertheless caused great headaches for the founding
fathers of QFT. They arise because of loop diagrams in QFT, whose
short-distance contributions involve divergent integrals. We thus call them {\em UV
  divergences}. But to actually get a divergence requires us to assume that the theory is valid on
arbitrarily short distance scales, way beyond those that we actually
probe in experiments. This seems overly arrogant and liable to result
in hubris. Indeed, it runs contrary to what
we have observed in all previous instances in physics, namely that
physical theories only ever have some limited region of
validity.\footnote{One day, of course, some bright spark might write down a
  theory of everything, in which case they would be quite justified in
  extrapolating in this way. But this can only happen once!}

\subsection{Effective field theory: na\"{\i}ve approach}
The point of departure for EFT is to humbly accept that any given
theory is likely to have some short-distance or UV cut-off, $\Lambda$, beyond
which it is invalid. We should not dare to extrapolate beyond this
cut-off. If we don't, then we will never encounter any UV divergences,
and so the problems that plagued the founding fathers of QFT seem to
have completely disappeared!

In its place, a new problem appears. The good side of insisting on
renormalizability (that is, a theory valid on all scales), was that it
necessarily restricted the dimensions of operators that can appear in
the action and hence implies that the theory has a finite number, $n$, of
parameters and hence is predictive, once we have made $n$
measurements. If we give up on renormalizability, but still write down
all operators consistent with the symmetry (if we don't, quantum
fluctuations will generate them anyway via the RG flow \dots), then we will have to
include infinitely many. (Proof: consider any operator that is
invariant under the symmetry; the $m$th power of the operator is also
invariant, for any $m \in \mathbb{Z}$.) If each of these operators has
an arbitrary coefficient, then we need to do infinitely many
measurements before we can start to make predictions. This is not a theory!

We find a way out of the impasse {\em \`{a} la} George Orwell, by
declaring that `all operators are equal, but some are more equal than
others'. How? Since we are interested in the physics at large-distance scales,
it may be that some operators are more important at
large-distances than others. This is indeed the case, and in fact it
turns out the usual QFT dimensional analysis gives us a measure of how
important operators are, relative to the kinetic term of the free theory (which governs
the size of typical fluctuations). 

Consider, as an example, relativistic scalar field theory in $D$
spacetime dimensions.\footnote{We will return to this example
  repeatedly in the sequel.} In units where $\hbar = c =1$, the action
is dimensionless and so the kinetic term, $(\partial \phi)^2$, has
energy dimension $D$. Since the 
derivatives have unit (energy) dimension, the field $\phi$ must have
 dimension $D/2-1$. An operator $\mathcal{O}_{p,q}$ made up of $p$ fields
and $q$ derivatives then has dimension $p(\frac{D}{2}-1)+q$ and appears in the
action as
\begin{gather}\label{eq:lag}
S \supset \int d^D x
\;\frac{g_{p,q}}{\Lambda^{p(\frac{D}{2}-1)+q-D}} \mathcal{O}_{p,q},
\end{gather}
where we have written the coupling in terms of the cut-off scale
$\Lambda$ and a dimension{\em less} coefficient, $g_{p,q}$. Now, we
see that if we consider a field configuration of energy $E$, the
contribution of the operator $\mathcal{O}_{p,q}$ to the action is, on
dimensional grounds,
given by 
\begin{gather}
\label{eq:power}
S \supset
\;g_{p,q} \left(\frac{E}{\Lambda}\right)^{p(\frac{D}{2}-1)+q-D} .
\end{gather}
If the exponent $p(\frac{D}{2}-1)+q-D > 0$, then the operator
becomes less and less important at energies below the cut-off and we
call it {\em irrelevant}. If $p(\frac{D}{2}-1)+q-D < 0$ then the
operator becomes more and more important at $E < \Lambda$ and we call
it {\em relevant}. If $p(\frac{D}{2}-1)+q-D = 0$ (which includes the kinetic
term itself), then the operator is equally important as the kinetic
term at low energies and we call it {\em marginal}.

Before going further, let us make two remarks. The first remark is
that in a non-relativistic theory, we will need to count dimensions of
space and time separately. The second remark is that our counting of
dimensions and our decision of which operators are (ir)relevant is
contingent on our singling out a particular term as `the' kinetic
term. This is a natural thing to do, since a kinetic term is present
in all dynamical theories and sets the scale for the typical size of
fluctuations in the theory. But there is no obvious definition of what
a `kinetic term' actually is and a given theory might 
have multiple kinetic term candidates. In such a case, one should
proceed by computing the dimensions of operators with respect to each
of these terms individually; it may turn out that different kinetic terms
dominate in different regimes of distance and time scales. 

Now let us return to the main thrust. We have discovered that, of the
infinitely many operators that we may write in the action, some are
more important than others at the large distance and time scales in
which we are interested. Can we use this to make a predictive theory?
The answer is no, strictly speaking. But we can use it to make a
theory which is almost as good, in that we can use it to make
predictions to an arbitrarily high degree of precision, provided that
we are willing to do enough donkey work.\footnote{Much the same is true, of
  course, when we do renormalizable QFT perturbatively; there, finite
  precision arises because we truncate the loop expansion.}

What we do is to write out the most general action, but including
operators only up to some finite dimension $\Delta$.\footnote{A
  renormalizable theory, then, corresponds to the special case with
  $\Delta = D$.} This truncated theory has only a finite number of
arbitrary coefficients so we can use it to make predictions, once we
have made enough measurements. (Clearly we will need to do more and
more measurements as we increase $D$ and herein lies some of the
aforementioned donkey work.) But we will not be able to make exact
predictions, because we have neglected operators in the theory whose
dimensions exceed $\Delta$. Comparing with \ref{eq:power}, we see that
in computing the action (or indeed any other observable), we have only
included contributions of $O
\left(\left(\frac{E}{\Lambda}\right)^{p(\frac{D}{2}-1)+q-D}\right)$
compared to the leading ones and so this is the accuracy of our
prediction. 

Four remarks are now in order. Firstly, we note that our predictions automatically
become arbitrarily accurate as we go to arbitrarily large distance
scales, {\em viz.} $E \rightarrow 0$. It is in this sense that we have
a theory for physics on large distance scales. Secondly, we note that
we can
improve the accuracy of our theory at fixed energy $E$ by truncating
at higher order in the operator expansion. To do this, one needs
to find all the invariant operators up to a given dimension (in
general, this number grows exponentially with the dimension), to
calculate the theory predictions including all these operators, and to
perform more measurements (at the higher accuracy) to fix the extra
parameters. This is a lot of donkey work. Thirdly, we remark that once
we hit energies $E$ of the order of the cut-off $\Lambda$, no amount
of donkey work is going to help us, because all neglected terms become
equally important. The expansion breaks down completely and so
the cut-off $\Lambda$ really does deserve its name. Finally, we remark
that we are not free to choose the value of $\Lambda$
arbitrarily. The predictions of the theory for experimental
observables depend, via (\ref{eq:power}), on $\Lambda$. And so we can
use measurements to determine the value of $\Lambda$ in a given
theory. 

\subsection{Effective field theory, {\em comme il le faut}}
So far, we implied that the way to do EFT is to impose a hard UV
cut-off on the theory, such that the UV divergences coming from loop
integrals in the theory do not appear. If we do this (as most lecture
notes, {\em \&c} do), then whilst we end up with a theory that is
manifestly finite, we also end up with a theory that is completely
useless for making predictions. The problem is that higher-dimension
operators give contributions that are suppressed when they appear in
tree-level Feynman diagrams, but not when they are inserted into
loops. It is easy to see schematically why this happens. At
tree-level, the only powers of $\Lambda$ that appear in amplitudes are
those coming from the denominators in (\ref{eq:lag}). So the presence
of higher-dimension operators always leads to suppression of
amplitudes by factors of $E/\Lambda <1$. But when we insert
higher-dimension operators into loops, we get additional powers of
$\Lambda$ in the numerators of amplitudes, coming from the fact that
we cut off the loop momenta at $\Lambda$. With enough loops, we can
always arrange for more powers of $\Lambda$ in the numerator than in
the denominator, meaning that the contributions of higher-dimension
operators will be unsuppressed. But then we are not at liberty to
simply truncate the operator expansion and ignore operators above a
certain dimension!

We can see the phenomenon explicitly using our
favourite example of scalar field theory. Consider 1-loop corrections
to the dimension-4 operator $\lambda \phi^4$. The EFT Lagrangian is
\begin{equation*}
\mathcal{L} = -\frac{1}{2}\phi(\pt^2 + m^2)\phi - \frac{1}{4!} \lambda \phi^4 \\
- \frac{1}{6!} \frac{c_6}{\Lambda^2} \phi^6 - \frac{1}{2 \cdot 4!} \frac{c_8}{\Lambda^4} \phi^4 (\pt \phi)^2 - \ldots,
\end{equation*}
where the dimensionless coefficients $c_6, c_8, \ldots$ are
$O(1)$. With momentum cut-off $\La$, we get loop diagrams of
similar size from {\em all} operators.\footnote{An apparently simple solution to this problem would be to use a lower cut-off $\La^\prime < \La$ for the loop integral. But doing so generates operators with derivatives of size $\frac{\pt}{\La^\prime}$ under the renormalization group flow, thereby reducing the regime of validity of the EFT as a whole to $p \lesssim \La^\prime$.} Indeed,
\begin{align}
\delta \lambda_{\text{1-loop}} &\supset \frac{c_6}{\La^2} \int^\La \frac{d^4 k}{(2\pi)^4} \frac{1}{k^2-m^2} \sim \frac{c_6}{\La^2} \frac{\La^2}{16 \pi^2} \sim \mathrm{O}(1), \\
\delta \lambda_{\text{1-loop}} &\supset \frac{c_8}{\La^4} \int^\La \frac{d^4 k}{(2\pi)^4} \frac{k^2}{k^2-m^2} \sim \frac{c_8}{\La^4} \frac{\La^4}{16 \pi^2} \sim \mathrm{O}(1), \; \mathit{\&c.}  
\end{align}
Thus we find that predictivity is lost using such a cut-off, since we need to consider loops containing all operators to calculate at any given order in the momentum expansion of the Lagrangian.

The solution to this problem is, in fact, very simple: we need to
replace the UV cut-off $\Lambda$ with a \emph{mass independent}
regulator, such as dimensional regularization.
Then, the only mass scales that can appear in the numerators of diagrams correspond to light masses or momenta, with the renormalization scale appearing only in logarithms. For the EFT of a scalar, for example,
\begin{align}
\frac{c_6 \mu^{2\epsilon}}{\Lambda^2} &\int \frac{d^{4-\epsilon} k}{(2\pi)^{4-\epsilon}} \frac{1}{k^2-m^2} \sim \frac{c_6}{\Lambda^2} \frac{m^2}{16 \pi^2} \frac{1}{\epsilon} - \frac{c_6}{\Lambda^2} \frac{m^2}{16 \pi^2} \ln( \frac{m^2}{\mu^2} ), \\
\frac{c_8 \mu^{2\epsilon}}{\Lambda^4} &\int \frac{d^{4-\epsilon} k}{(2\pi)^{4-\epsilon}} \frac{k^2}{k^2-m^2} \sim \frac{c_8}{\Lambda^4} \frac{m^4}{16 \pi^2} \frac{1}{\epsilon} - \frac{c_8}{\Lambda^4} \frac{m^4}{16 \pi^2} \ln( \frac{m^2}{\mu^2} ), \; \mathit{\&c,} 
\end{align}
where $\mu$ is the renormalization scale.

A mass independent scheme thus preserves the original momentum
expansion: contributions from higher dimension operators are
suppressed, even in loops. If we consider all operators up to
dimension $\Delta$, we are guaranteed a result accurate to $O \left(
  (E/\La) ^{\Delta-4} \right)$, where $E$ is the energy scale of the
process, at any loop order.

\subsection{Topsy-turvy EFT}
We can motivate the EFT idea in a completely different way, by showing
that starting from a renormalizable QFT at high energies, the low
energy theory is equivalent to an EFT. 

Suppose, for example, that we start with the renormalizable SM,
and consider only energies and momenta well below the weak scale,
$\sim 10^2$ GeV. We can never produce $W$, $Z$, or $h$ bosons on-shell
and so we can simply do the path integral with respect to these
fields (we `integrate them out', to use the vernacular). We will be left
with a path integral for the light fields, but with a complicated
lagrangian that is non-local in space and time (where non-local means
that the lagrangian density cannot be written in terms of the fields and a finite
numbers of derivatives thereof, evaluated at a single spacetime point). But since we are only interested in low energies
and momenta, we can expand in powers of the spacetime derivatives (and
the fields)
to obtain an infinite series of {\em local} lagrangian operators, which become
less and less important as we go down in (energy-)momentum. 

At tree-level,
this procedure just corresponds to replacing the fields using their classical
equations of motion, and expanding $\frac{-1}{q^2-m_W^2} =
\frac{1}{m_W^2} + \frac{q^2}{m_W^4} + \dots$.
Note that the LHS of this expression is non-local, while each term in
the sum on the RHS is local.
It is already clear that
our expansion breaks down for momenta comparable to $m_W$, so that the
theory is naturally equipped with a cut-off scale $m_W$.

In particular, the leading operator we get by the above process will
be the 4-fermion operator in Fermi's theory of beta decay, with
coefficient $\sim \frac{1}{m_W^2}$. By measuring the decay constant,
$G_F$, we are able to estimate the cut-off $m_W$. All of this is in
accord with what we discussed above.
\subsection{The scourge of relevant operators}
Now is the time for us to acknowledge the presence of an elephant in the room. In
renormalizable QFT, the problems come from irrelevant operators. They
are non-renormalizable and lead to uncontrollable divergences in loop
diagrams. But in EFT, irrelevant operators are completely
benign. There are no divergences, and instead the irrelevant
operators are, well, irrelevant. Or at least, {\em largely}
irrelevant, in that they give small corrections to physics at energy
scales well below the cut-off. 

In EFT, the problems come rather from relevant operators. These become
increasingly important at low energies, and indeed (\ref{eq:power})
shows that they dominate the physics. But this invalidates our
assumption that the physics is dominated by the kinetic term, and so
invalidates our operator expansion. All we can say is that the physics
of the system at low energies is likely to be completely different
from that `predicted' by the original EFT. 

To examine this in more detail, let us start with a relatively trivial
case. Consider scalar field theory in 4-d. The symmetries allow a
mass term $\propto \phi^2$ in the lagangian. This has dimension 2
(meaning that we can write its coefficient as $g_{2,0}\Lambda^2$, with
$g_{2,0}$ being dimensionless) and it gives contributions of size
$g_{2,0}\Lambda^2/E^2$ relative to the kinetic term. There are then
two possibilities. Either $g_{2,0} \gtrsim 1$, in which case this term
always dominates the kinetic term. We should redo our scaling
arguments above, taking $\phi^2$ to be the dominant term at low
energies. If we do, we will find that all other operators are
irrelevant. At low energies therefore, the dynamics is dominated by
the term $\phi^2$. Classically, we find that $\phi = 0$ and there is
no dynamics at all. We obtain a consistent theory of nothing! The
alternative is that $g_{2,0} \ll 1$, in which case there is a regime
of energies in which the kinetic term dominates and our EFT is
valid. But then the question arises of how we can end up with a theory
in which $g_{2,0} \ll 1$. Indeed, starting from a generic short-distance
theory of dynamics at the scale $\Lambda$ we will invariably end up
with $g_{2,0} \sim 1$ in the low energy EFT. Again, a simple example
suffices to illustrate the general point: consider a theory with two
scalar fields $\phi$ and $\Phi$, where $\phi$ is assumed light
compared to $\Phi$, which has mass $M$. If we
integrate out the field $\Phi$ to obtain the low-energy EFT for the
light scalar $\phi$, we will find that loops of $\Phi$ give
corrections to the mass of $\phi$ of order $M$. 

Thus, to end up with a small mass for $\phi$, we need to delicately
arrange the tree-level and loop contributions (which correspond to
physics on differing length scales) in order to obtain a cancellation
in the resulting value. This is called an {\em unnatural
  fine-tuning}. 

Unfortunately, this issue is not just an academic one: the Standard
Model of Particle Physics features just such a scalar field (the
Higgs field) and it is a mystery to us why its mass is so light
compared to the short-distance theory that gives rise to the SM.

Finally, note that it is quite possible to have relevant operators
that are not mass terms, but rather correspond to interactions. Are
these bad too? They certainly are, because they represent interactions
that become arbitrarily strong at low energy. Perturbation theory thus
breaks down completely. All it is safe to say is that the
degrees of freedom and symmetries that we assumed in formulating our
EFT are completely unsuitable for describing the physical system at
low energies.

You already know a good example of this, namely QCD, where the
coupling is marginal at tree-level, but acquires an anomalous
dimension and becomes relevant at one-loop. The coupling thus becomes
strong at low energies, and the low energy degrees of freedom (mesons
and baryons) are completely unlike the quarks and gluons of QCD.

\section{First example: The Standard Model and beyond}
We have already described how Fermi's theory of beta decay is just a
low-energy EFT description of a more complete, short-distance theory,
{\em viz.} the SM. Now that we know about EFT and how it works, it
seems reasonable to suppose that every QFT we have to hand is really
just a low-energy EFT description of some more fundamental theory. Let
us suppose that the SM itself is just an
effective, low-energy description of some more complete BSM theory,
and see what the consequences may be.

Following the rules above, the fields and the (gauge) symmetries of
the EFT should be exactly the same
as in the SM, but we should no longer insist on renormalizability. For
operators up to dimension 4, we simply recover the SM. But at
dimensions higher than 4, we obtain new operators, with new physical
effects. As a striking example of these effects, we expect that the accidental
baryon and lepton number symmetries of the SM will be violated at
some order in the expansion, and that protons will decay.\footnote{Let us
  hope that we can finish the lecture before they do so!} 

We don't know what the BSM theory actually is yet, and so when we
write down the EFT, we should allow the coefficients of the
operators in the expansion to be arbitrary. 
While we don't know the actual values of the
coefficients, we can estimate their size using dimensional
analysis, since we expect the expansion to break down at energies of
order the cut-off, $\Lambda$. So the natural size of coefficients is typically
just an $O(1)$ number in units of $\Lambda$, which is precisely how we
wrote them above.
\subsection{Mathematical interlude on vector spaces}
Now we wish to write down the most general set of operators up to a
given dimension. Before doing so, it is useful to notice 
that the operators of a given dimension form a
vector space, $V$, and so we can simplify things by choosing a basis for
this space. 
This is not so straightforward as it sounds (and indeed,
disputes about it still erupt in the literature from time to time), because of
equivalences between operators. In particular, any two operators that
are equal up to a total derivative may be considered equal (since they
give the same contribution at any order in perturbation theory), as may
operators that differ by terms that vanish when the equations of
motion hold, because such pieces give contributions to $S$-matrix
elements that are non-vanishing only at higher orders (see, {\em e.\ g.},
\cite{Arzt:1993gz}).

For a simple example \cite{Einhorn:2013kja}, consider a scalar field theory, in
which we allow only operators that are even in $\phi$ and set the mass
term to zero, for simplicity. The lagrangian at dimension 4 is then
\begin{gather}
\mathcal{L} = \frac{1}{2}(\partial \phi)^2 - \frac{\lambda}{4} \phi^4 .
\end{gather}
At dimension 6, three operators present themselves, namely $\phi^6,
(\partial^2 \phi)^2,$ and $\phi^2 (\partial \phi)^2$. Only one of
these is independent. Indeed, we have that
\begin{gather}
(\partial^2 \phi)^2 - \lambda^2 \phi^6 = [(\partial^2 \phi) - \lambda \phi^3][(\partial^2 \phi) + \lambda \phi^3]
\end{gather}
and the second term on the right vanishes when the equations of motion
hold. 

Similarly, integrating by parts we have that
\begin{gather}
\phi^2 (\partial \phi)^2 = -\phi \partial_\mu (\phi^2 \partial^\mu
\phi) = -\phi^3 \partial^2 \phi -2 \phi^2 (\partial \phi)^2
\end{gather}
which implies that 
\begin{gather}
3\phi^2 (\partial \phi)^2 = -\phi^3 \partial^2 \phi 
\end{gather}
and thus that
\begin{gather}
3\phi^2 (\partial \phi)^2 - \lambda \phi^6 = -\phi^3 (\partial^2 \phi
+\lambda \phi^3).
\end{gather}
Thus, we see that $\phi^2 (\partial \phi)^2 \sim \lambda^2 \phi^6$ and
$3\phi^2 (\partial \phi)^2 \sim \lambda \phi^6$, where $\sim$ denotes
equivalence.

The best way to deal with these equivalences is as follows. Write each
equivalence in the form $A=0$, where $A$ is an operator in the vector space $V$ and let $U
\subset V$ be the linear span (i.e. all linear combinations) of the
$A$s. Now we form equivalence classes in $V$ by identifying any two
operators that differ by an operator in $U$. So, for example, if $B =
C + A$, then we regard $B$ and $C$ as equivalent operators and write
$[B] = [C]$, where $[B]$ denotes the class containing $B$. In doing
so, we form the quotient space, $V/U$, of equivalence classes. This is
itself a vector space, with zero vector $[0]=U$, where $0$ is the zero
vector in $V$. 

As you can see, identifying a true basis of operators is not
easy, even for the simple example of scalar field theory. Fortunately, there will soon be a computer
program that will do it for you \cite{sutherland} (at least for a
theories like the SM whose symmetries only include factors of $SU(N)$
and $U(1)$).

It is common in the literature to see a further subdivision of
operators (or, rather, equivalence classes of operators) into those
that can be generated at $n$-loop level in a renormalizable UV
completion, where $n \in \{0,1,2,3,\dots \}$. The rationale for doing
this is that, if the new physics couplings are of $O(1)$, then each
additional loop leads to suppression factor of $\sim 4\pi$, lowering
the scale of new physics (i.e. the cut-off) that is required to
generate a contribution of a given size. 

What the literature does not tell you, sadly, is that the classes of operators generated at a given loop level do {\em
  not} form a vector subspace, in general.\footnote{For a
  counterexample, consider scalar field theory in $d=6$. The scalar
  field $\phi$ has dimension 2 and so the $\phi^3$ interaction is
  marginal. At dimension 8, the only class of operators is
  $[\phi^4]$. The operator $+[\phi^4]$ can be generated at tree level,
but the operator $-[\phi^4]$ can only be generated at one-loop
level. So the tree level operators cannot form a vector subspace. For
more details, see \cite{sutherland}.} Thus, it is
meaningless to set experimental limits on the scale of new physics by
taking an arbitrary linear combination of, say, classes of operators
that can be generated at tree level: the resulting class of operators
is not necessarily tree-level generated. This doesn't stop people
doing it though!
\subsection{Back to the SM}
Getting back to the SM, let's start by reminding ourselves of the form
of the lagrangian. Recall that the SM is a gauge
theory with gauge symmetry $SU(3) \times SU(2) \times
U(1)$, together with matter fields comprising 15 Weyl fermions and one
complex scalar, carrying irreps of $SU(3) \times SU(2) \times
U(1)$. The fermions consist of 3 copies (the different families or
flavours or generations) of 5 fields, $\psi \in \{q,u^c, d^c, l, e^c\}$,
carrying reps of $SU(3) \times SU(2) \times
U(1)$ as listed in Table~\ref{tab:smreps}. The 
scalar field, $H$, carries the $(1,2,-\frac{1}{2})$ rep of $SU(3) \times SU(2) \times
U(1)$.
\begingroup
\begin{table*}[ht]
\begin{center}
\begin{tabular}{c  c  c  c  }
\hline
\hline
Field &  $SU(3)_c $ & $SU(2)_L $   & $U(1)_Y $  \\
\hline
$q $ &3 &2 &$+\frac{1}{6}$ \\
$u^c$ & $\overline{3}$& 1& $-\frac{2}{3}$\\
$d^c$ & $\overline{3}$& 1& $+\frac{1}{3}$\\
$l $ &1 &2 & $-\frac{1}{2}$ \\
$e^c$ &1 & 1& $+1$\\
\hline
\end{tabular}
\end{center}
\caption{Fermion fields of the SM and their $SU(3)\times SU(2) \times
U(1)$ representations.\label{tab:smreps}}
\end{table*}
\endgroup

The lagrangian can be written
on a single line (just!). It is, schematically,
\begin{gather}
\mathcal{L} = i \overline{\psi}_i \overline{\sigma}^\mu  D_\mu \psi_i
-\frac{1}{4} F^a_{\mu \nu} F^{a \mu \nu} + \lambda^{ij} \psi_i \psi_j
H^{(c)} + \mathrm{h.\ c.}+ |D_\mu H|^2  - V(H),
\end{gather}
where $i,j$ label the different families and $a$ labels the different
gauge fields. There are
5 fermion irreps $\psi \in \{q,u^c,d^c,l,e^c\}$, with 3 copies of
each, corresponding to the 3 families. There are really 12 gauge
fields: 8 in an adjoint of $SU(3)$, 3 in an adjoint of $SU(2)$, and 1
for $U(1)$. The covariant derivative $D_\mu$
contains the gauge couplings $g_s, \, g,$ and $g^\prime$, with the gauge
group generators in the appropriate reps.
The fermion kinetic terms (but not the Yukawa couplings) are invariant under a $U(3)^5$ global
symmetry. 
The Yukawa interactions can be written more explicitly as
\begin{gather} \label{eq:lyuk}
\mathcal{L} = \lambda^u q H^c u^c + \lambda^d qH d^c + \lambda^e l H e^c +
\mathrm{h.\ c.}
\end{gather}
The $\lambda^i$ are 3 $3\times 3$ complex matrices (in family space).

The Higgs potential is given by
\begin{gather}
V(H) = \mu^2 H^\dagger H + \lambda (H^\dagger H)^2 .
\end{gather}

Ugly or not, the renormalizable SM does an implausibly good job of describing the
data, reaching the per mille level in individual measurements and with
an overall fit (to hundreds of measurements) that cannot be denied:
the SM is undoubtedly correct, at least in the regime in which we are currently
probing it (see \cite{Gripaios:2015gxa} for more details, at a similar
level to these lectures). What does this imply, if the SM is really just an EFT,
with a cut-off $\La$? Since the operators with dimension up to 4
already do an excellent job of describing the measurements (which are
themselves very precise), we must conclude that the effects of
higher-dimension operators are very small. In other words, $\La$ must
be very large. How large? Well, each experimental measurement that
agrees with the SM predictions can be translated into a rough lower
bound on $\La$, once we make the reasonable assumption that the
dimensionless coefficients are of order 1. In this way, we obtain some
very stringent bounds on $\Lambda$, reaching up to $10^{15}$ GeV or
so! This is way beyond the reach of the LHC.

\subsection{Accidental symmetries and proton decay}
One miracle of the SM is that it has accidental symmetries. These
are symmetries of the lagrangian that are not put in by {\em fiat},
but arise accidentally from the field content and {\em other} symmetry restrictions,
and the insistence on renormalizability. Once we allow operators with
higher dimensions in the EFT, we will find that these accidental
symmetries get broken, with sometimes spectacular consequences for physics.

A simple example of an accidental symmetry is parity in QED.
The most general, Lorentz-invariant, renormalizable lagrangian for electromagnetism coupled to a Dirac
fermion $\Psi$ may be written as
\begin{gather}
\mathcal{L} = -\frac{1}{4} F_{\mu \nu} F^{\mu \nu} + ia F_{\mu \nu} \tilde{F}^{\mu \nu} +
i\overline{\Psi}\slashed{D} \Psi + \overline{\Psi} (m + i\gamma^5 m_5) \Psi,
\end{gather}
where both the term involving $\tilde{F}^{\mu \nu} \equiv \epsilon
^{\mu \nu \sigma \rho}F_{\sigma \rho}$ and the term involving
$\gamma^5$ na\"{\i}vely violate parity. However, the former term is a
total derivative and so does not contribute to physics at
any order in perturbation theory. The latter term can be removed by a chiral rotation $\psi
\rightarrow  e^{i\alpha \gamma^5}\psi$ to leave a parity-invariant
theory with fermion mass $\sqrt{m^2 + m_5^2}$. So we find that the
lagrangian is invariant under parity, even though we did not require
this in the first place. The same is true of charge conjugation
symmetry. Note that if we had not insisted on renormalizability, we
could write dimension-six terms like $\overline{\Psi} \gamma^\mu
\gamma^5 \Psi \overline{\Psi} \gamma_\mu \Psi$, which do violate
parity. 

As we already alluded to above, the SM lagrangian is accidentally
invariant under a $U(1)_B$ baryon number symmetry (an overall rephasing
of all quarks)
and three $U(1)$ lepton number symmetries, corresponding to individual
rephasings of the three different lepton families (which contains an
overall lepton number symmetry $U(1)_L$ as the diagonal subgroup).
Either $U(1)_B$ or $U(1)_L$  symmetry, together with Lorentz invariance, prevents the proton from
decaying. Indeed, a putative final state must (by Lorentz invariance,
which implies the fermion number is conserved mod 2) contain an odd number of fermions lighter than the proton. The only such states
carry lepton number but not baryon number, whereas the proton carries
baryon number but not lepton number.

Again, once we allow higher dimension operators, we will find that
lepton and baryon number are violated (by operators of dimension five
or six, respectively), meaning that the proton can decay. Similarly, generic theories of physics BSM will
violate them and hence will be subject to strong constraints.

There is another interesting accidental symmetry of the SM, which is
only approximate. This is called {\em custodial symmetry}. Consider the Higgs sector. The Higgs is a
complex $SU(2)$ doublet, and so there are four real fields. The
kinetic terms therefore have an $O(4)$ symmetry. Let us now consider
how this symmetry gets broken when we switch on the various
couplings. 

One of the miracles of group theory is that the Lie algebra of the
group $O(4)$ is the same as that of the group $SU(2) \times SU(2)$. So
the Higgs fields can be thought of as carrying 2 $SU(2)$ symmetries,
rather than the single $SU(2)_L$ of the standard model. It is usual to
call the other symmetry $SU(2)_R$, so the Higgs carries a $(2,2)$ rep of
$SU(2)_L \times SU(2)_R$.
Now, when we switch on the
$SU(2)_L$ gauge coupling $g$, we still have global symmetry $SU(2)_L$
(because the gauge symmetry includes constant gauge transformations,
which are the same as the global ones) and we still have global
symmetry $SU(2)_R$, because this factor is independent of $SU(2)_L$.
So the full $SU(2)_L \times SU(2)_R$ remains unbroken.

What is more, this $SU(2)_L \times SU(2)_R$ is also unbroken when we
switch on the
Higgs potential, because $V(H)$ is only a function of $|H|^2 = h_1^2
+h_2^2 +h_3^2 +h_4^2$, which is manifestly invariant under $O(4)$.

The Yukawa couplings do break $SU(2)_L \times SU(2)_R$,\footnote{A
  technical point: if $\lambda^u = \lambda^d$, then we can group $u^c$
and $d^c$ into an $SU(2)_R$ doublet, and $SU(2)_L
\times SU(2)_R$ is restored.} as does the
coupling to the $Z$ (which couples to the combination $T^3_L + T^3_R$).
So the correct statement is that the SM is invariant under $SU(2)_L
\times SU(2)_R$ in the limit that $\lambda^u = \lambda^d, g^\prime =0$. 

When the Higgs gets a VEV, the $SU(2)_L \times SU(2)_R$ is broken to
the diagonal $SU(2)_V$ combination of the 2 original $SU(2)$s. This
approximate symmetry implies a relation between $m_W$ and
$m_Z$ that holds automatically in the SM, but does not hold in generic
theories BSM. Again, see \cite{Gripaios:2015gxa} for more details.

\subsection{Beyond the SM - Effective field theory}
Now let's reconsider the SM from the EFT viewpoint, cataloguing the
operators of increasing dimension and describing their effects in turn.
\subsection{$D=0$: the cosmological constant}
We have avoided mentioning it up to now, but clearly a constant term
(which has dimension 0) is
consistent with the symmetries of the SM. It has no effect until the
SM is coupled to gravity, whereupon it causes the Universe to
accelerate. On the one hand, this looks like good news, because the
Universe is observed to accelerate. On the other hand, this is bad
news because our estimate of the size of this operator coefficient
(the operator is 1) is $\Lambda^4$, while the observed energy density is around $(10^{-3}~
\mathrm{eV})^4$. But the cut-off of the SM had better not be $10^{-3}~
\mathrm{eV}$, because if it were then we could certainly not use it to
make predictions at LHC energies of several TeV. So either dynamics or
a tuning makes the constant small. If we consider the Planck scale to be a real physical cut-off, then we need to tune at the level of 1 part in $10^{120}$. It
is fair to say, that despite $O(10^{120})$ papers having been written
on the subject, no satisfactory
dynamical solution has been suggested hitherto. An alternative is to argue that
we live in a multiverse in which the constant takes many different
values in different corners, and we happen to live in one which is
conducive to life. Indeed, it has been argued \cite{Weinberg:1988cp} that if the constant
were much larger and positive, structure could never form, while if it
were too large and negative, the Universe would re-collapse before
life could appear. The flavour-of-the-month as regards how the
multiverse itself arises is by a process of eternal inflation in
string theory.
\subsection{$D=2$: the Higgs mass parameter}
The only other relevant operator in the SM is the Higgs mass
parameter, which sets the weak scale. As above, the natural size for this is
$\Lambda$. But we measure $v \sim 10^2$ GeV, leaving us with 2
options:
either the natural cut-off of the SM is not far above the weak scale
(in which case we can hope to see evidence for this, in the form of new
physics, at the LHC) or the cut-off is much larger, and the weak scale
is tuned, perhaps once again by anthropics. 
\subsection{$D=4$: marginal operators}
We have discussed these already in the context of the renormalizable
SM, and there is nothing to add here. 
\subsection{$D=5$: neutrino masses and mixings}
Now things get more interesting. There is precisely one operator at
$D=5$, namely $\frac{\lambda^{ll}}{\Lambda} (lH^c)^2$, where
$\lambda^{ll}$ is a dimensionless $3\times 3$
matrix in flavour space. Note that this operator violates the
individual and total lepton numbers; moreover, it gives masses to
neutrinos after EWSB, just as we observe. So, one might argue that it is no surprise that neutrino masses
have been observed, since they represent the {\em leading} deviation from
the SM, in terms of the operator expansion.
Given the observed $
10^{-3} ~\mathrm{eV}^2$ mass-squared
differences of the neutrinos, we estimate $\Lambda \sim 10^{14}$
GeV. Thus, one could argue
that while neutrino masses are undeniably, as one so often hears, evidence for physics BSM,
they are also evidence that the SM is valid up to energy scales that
are way, way beyond the reach of conceivable future colliders.

Even so, it is worthwhile to consider what theory might replace the
EFT at $\Lambda$ to give a UV completion, extending the regime of
validity. One extremely simple possibility is to add to
the SM a new fermion, $\nu^c$, that is a singlet under $SU(3)\times SU(2) \times
U(1)$. In fact we need at least 2 of these to generate the two
observed neutrino mass-squared differences, and it seems plausible
that there are 3 -- one for each SM family.

We may then replace the $D=5$ operator with the renormalizable Yukawa
term $\lambda^\nu l H^c \nu^c$ (which is a Dirac mass term for
neutrinos after EWSB), along with the Majorana mass term $m^\nu \nu^c
\nu^c$. By making $m^\nu$ large, light neutrino masses can be
generated, even with $\lambda^\nu = O(1)$. This is the so-called `see-saw' mechanism, about which
you may have heard. 
\subsection{$D=6$: trouble at t'mill}
Once we get to $D=6$, a whole slew of operators appear. These include
operators that violate baryon and lepton number, such as $\frac{qqql}{\Lambda^2}
$ and $\frac{u^c u^c d^c e^c}{\Lambda^2}$ and which cause the
proton to decay via $p \rightarrow e^+ \pi^0$. We can estimate a lower bound on $\Lambda$ from the experimental bounds on the proton
lifetime, $\tau_p > 10^{33}$ yr, as follows. The decay rate (which comes from the amplitude
squared) is proportional to $\frac{1}{\Lambda^4}$ and the remaining
dimensions must be supplied by phase space, giving a factor of
$m_p^5$. Plugging in the numbers, we get $\Lambda > 10^{15}$
GeV. Again, the implication is that new physics either respects baryon
or lepton number, or is a long way away.

There are also operators that give corrections to flavour-changing
processes that are highly suppressed in the SM, because of the GIM
mechanism, and to which experiments are therefore unusually sensitive. As an example, the operator $(s^c d) (d^c s)/\Lambda^2$ contributes to Kaon mixing
and measurements of $\Delta m_K$ and $\epsilon_K$ yield a bound of
$\Lambda > 10^5$ TeV. 
\subsection{Two pitfalls}
\begin{figure}
\centering
\begin{tabular}{c c}

\begin{fmffile}{downedm_feynmp}
\begin{fmfgraph*}(100,100)
\fmfleft{din}
\fmfright{dout}
\fmftop{gamma}
\fmf{fermion}{din,v1}
\fmf{fermion,label=$u$,tension=0.5}{v1,v2}
\fmf{fermion}{v2,dout}
\fmf{photon,left=0.3,tension=0.5}{v1,vblob,v2}
\fmf{photon}{gamma,vblob}
\fmfblob{10}{vblob}
\fmflabel{$d$}{din}
\fmflabel{$d$}{dout}
\fmflabel{$\gamma$}{gamma}
\fmffreeze
\fmfiv{l=$W^-$,l.a=-20,l.d=.2w}{vloc(__vblob)}
\fmfiv{l=$W^+$,l.a=200,l.d=.2w}{vloc(__vblob)}
\end{fmfgraph*}
\end{fmffile}
\hspace{15mm}
&
\begin{fmffile}{upedm_feynmp}
\begin{fmfgraph*}(100,100)
\fmfleft{din}
\fmfright{dout}
\fmftop{gamma}
\fmf{fermion}{din,v1}
\fmf{fermion,label=$d$,tension=0.5}{v1,v2}
\fmf{fermion}{v2,dout}
\fmf{photon,left=0.3,tension=0.5}{v1,vblob,v2}
\fmf{photon}{gamma,vblob}
\fmfblob{10}{vblob}
\fmflabel{$u$}{din}
\fmflabel{$u$}{dout}
\fmflabel{$\gamma$}{gamma}
\fmffreeze
\fmfiv{l=$W^+$,l.a=-20,l.d=.2w}{vloc(__vblob)}
\fmfiv{l=$W^-$,l.a=200,l.d=.2w}{vloc(__vblob)}
\end{fmfgraph*}
\end{fmffile}

\end{tabular}

\caption{One-loop contributions of $\mathcal{O}_W$ (shaded blob) to the neutron EDM.\label{fig:nedm}}
\end{figure}
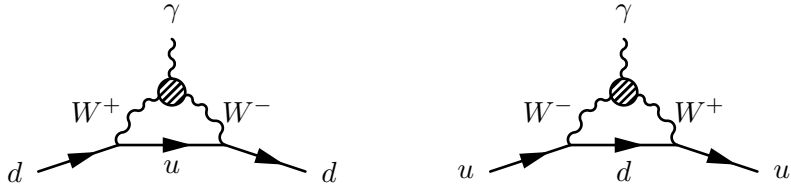

The SM affords a wonderful example of what goes wrong if one doesn't
regularize using a mass-independent scheme. Consider the dimension 6
operator $\mathcal{O}_W \propto \frac{i \epsilon^{abc}}{3!}
{W^a}^\mu_\nu {W^b}^\nu_\lambda \tilde{W}^c{}^\lambda_\mu$. This
operator violates $CP$ and thus may be relevant for baryogenesis, so
it is of interest to ask what the bound on its coefficient is. Now,
the operator $\mathcal{O}_W$ contributes to the electric dipole moment
of the neutron at one-loop, via the diagrams shown in Figure
\ref{fig:nedm}. Five sets of authors attempted this calculation in the
literature, obtaining five different results, mostly because the
authors were using a variety of regularization schemes. One set of
authors even showed that essentially any answer could be obtained by a
suitable choice of regularization! We know, of course, that only
results obtained using a mass-independent regulator are reliable.
In fact, this historical example affords us yet another illustration
of a classic pitfall. Since the SM $SU(2)\times U(1)$ gauge invariance is broken in the
vacuum, some authors have tried to argue that the correct way to write
the EFT expansion is in terms of operators that respect only the
unbroken subgroup of electromagnetism. In this example, one can write
not only $\mathcal{O}_W$, but also an arbitrary superposition of the two
electromagnetic invariants $\mathcal{O}_Z\equiv {W^+}^\mu_\nu {W^-}^\nu_\lambda \tilde{Z}^{\lambda}_\mu$ and
$\mathcal{O}_\gamma\equiv {W^+}^\mu_\nu {W^-}^\nu_\lambda
\tilde{F}^{\lambda}_\mu$. But it is easy to show that if the
coefficients of $\mathcal{O}_Z$ and $\mathcal{O}_\gamma$ are proportional to $\frac{1}{\La^{\prime 2}}$, then the
real cut-off of the EFT is not $\La^\prime$, but rather is $\sqrt{v
  \La^\prime}$. This is completely obvious if we work in a manifestly
$SU(2)\times U(1)$-invariant formalism, where the same physics can be
described by including the dimension {\em eight} operator $ H^\dagger {W}^\mu_\nu {W}^\nu_\lambda H \tilde{B}^\lambda_\mu $. See
\cite{Gripaios:2013lea} for details.
\section{Second example: Non-linear sigma models and the composite
  Higgs}
We have already argued that there is a basic problem with our
canonical example of scalar field theory as an EFT: it contains a
relevant operator, $\phi^2$, requiring either an unnatural fine tuning
of the parameters, or a breakdown of the EFT at low energies.

It turns out that it is possible to forbid this operator, and make a
consistent EFT of scalar fields by means of additional symmetries,
albeit in a non-trivial way. The trick is to make the scalar field a
Goldstone boson. 

You have probably encountered Goldstone bosons before in QFT in the
context of `spontaneous symmetry breaking'. This is a bad misnomer,
because if the symmetry really were broken, we could not use it to
forbid operators (like the mass term) in the lagrangian. It is better
to say that the symmetry is {\em non-linearly realized} in the vacuum.
Let's do it properly (see \cite{Preskill:1990fr} for more details).

Along the way, I'll illustrate the general results in the context of a
specific example, called the minimal composite Higgs model (MCHM)
\cite{Agashe:2004rs}.\footnote{For less minimal models, see
\cite{Gripaios:2009pe,Mrazek:2011iu}.} This is
one of the leading candidates for solving the electroweak hierarchy
problem. For more details, see \cite{Contino:2010rs,Panico:2015jxa}.

A general EFT theory of Goldstone bosons is called a {\em non-linear
  sigma model}. We suppose that there is a physical system with
dynamics invariant under a 
continous (Lie) symmetry group $G$, but such that the ground state is invariant only
under a proper subgroup $H \subset G$. Thus, if we act with an element
$h \in H$ on the ground state, we get it back again. But if we act
with a $U \in G$ but $U \notin H$, then we must obtain a different state. But this
state must also be a ground state, because the dynamics is invariant
under $G$. Thus the theory has a space of degenerate, inequivalent
ground states, much like the bottom of the Mexican hat in the
potential for the SM Higgs field.

In the MCHM, $G=SO(5)$ and $H=SO(4)$. $SO(n)$ is the group of $n
\times n$ orthogonal matrices with unit determinant. Its Lie algebra
is the vector space of $n \times n$ traceless, imaginary, Hermitian
matrices. There are $\frac{n(n-1)}{2}$ such matrices and so $SO(5)$ is
10 dimensional and $SO(4)$ is 6 dimensional.

\subsection{The coset space $G/H$ of inequivalent ground states}
How can we parameterise the space of ground states? Start with some
ground state, $\Phi_0$, pick two elements $U$ and $U^\prime$ of $G$
and consider the states $U\Phi_0$ and $U^\prime \Phi_0$. Clearly these
will be the same state if we can write $U^\prime = Uh$, with $h \in H$, since we know
that $h\Phi_0 = \Phi_0$. At this point it is useful to define an
equivalence relation\footnote{A relation $\sim$ between pairs of elements of
  a set $\{a,b,c,\dots\}$ is called an equivalence relation if (i) $a
  \sim a$, (ii) $a \sim b \implies b \sim a$, and (iii) $a \sim b$ and
  $b \sim c\implies a \sim c$ for all elements. It then follows that the sets of
elements that are equivalent to each other, called the equivalence classes, partition the
original set.}   by $U \sim U^\prime$ if $\exists h
\in H$ s.\ t. $ U^\prime = Uh$. The equivalence classes are
called the {\em left cosets} of $H$ in $G$, and there is one of them
  for every inequivalent ground state.

Now we can try to parameterize the space of cosets. A nice way to do
so is to choose an orthonormal basis $\{T^{\tilde{a}}, X^a\}$ for the Lie algebra of $G$,
such that $\{T^{\tilde{a}}\}$ are a basis for the Lie algebra of $H$. We may then parameterise
the cosets (and hence the vacua) by $U = e^{i\phi^a X^a}$.

In the MCHM, a suitable basis for $\{T^{\tilde{a}}\}$ is any set of
linearly independent traceless, imaginary, Hermitian
matrices with zeros in the fifth row and column. A particularly
convenient choice is
\begin{align}
T^{\tilde{a}_L} & = − \frac{i}{2}\left[ \frac{1}{2}
                \epsilon^{\tilde{a}bc}(\delta^b_i \delta^c_j -
                \delta^b_j \delta^c_i)+ (\delta^{\tilde{a}}_i \delta^4_j -
                \delta^{\tilde{a}}_j \delta^4_i)\right], \\
T^{\tilde{a}_R} & = − \frac{i}{2}\left[ \frac{1}{2}
                \epsilon^{\tilde{a}bc}(\delta^b_i \delta^c_j -
                \delta^b_j \delta^c_i)- (\delta^{\tilde{a}}_i \delta^4_j - \delta^{\tilde{a}}_j \delta^4_i)\right] ,􏰆􏰂􏰇
\end{align}
where $\tilde a,b,c \in \{1,2,3\}$. This choice is convenient, because when we work out the Lie brackets,
we find that the $T^{\tilde{a}_L}$ and $T^{\tilde{a}_R}$ form two independent copies of the
$SU(2)$ algebra. We thus learn that, at least at the level of the Lie
algebra, the group $SO(4)$ is equivalent to $SU(2) \times SU(2)$. We
say that they are {\em locally isomorphic}. This is important because,
as described above,
the SM has an approximate
accidental custodial symmetry $SU(2) \times SU(2)$, which we would
like to build in to any theory beyond the SM.

A suitable basis for $\{X^a\}$ are the matrices
\begin{gather}
X^a = - \frac{i}{\sqrt{2}}\left[ (\delta^a_i \delta^5_j -
                \delta^a_j \delta^5_i)\right]
\end{gather}

Note that there are 10-6=4 linearly independent matrices and that this
is also the number of scalar fields $\phi^a$ in the theory.

Why are we making such a big effort to parameterize the inequivalent ground
states of the theory? Suppose we now promote the parameters $\phi^a$
to spacetime fields $\phi^a (x)$. $\phi^a (x)$ = constant corresponds
to a ground state, but by making $\phi^a (x)$ vary arbitrarily slowly
in spacetime, we obtain an excitation of the theory that is arbitrarily close to the
ground state, and hence has abritrarily small energy. We can now try
to build an EFT for these low-energy excitations. 

How do we build the EFT? Clearly the appropriate degrees of freedom
are the fields $\phi^a(x)$ and the appropriate symmetry is $G$, but
how does it act on the fields $\phi^a(x)$? Under a $G$ transformation
with $\Omega \in G$, we know that a ground state $\Phi$ transforms to
$\Omega \Phi$.
But every ground state $\Phi$ can be written as $U\Phi_0$ with $U =
e^{i\phi^a X^a}$.
Thus we have that $e^{i\phi^a X^a}\Phi_0 \mapsto e^{i\phi^{\prime a}
  X^a} \Phi_0\equiv \Omega e^{i\phi^a X^a} \Phi_0$. Now, here we must
be careful. It is tempting to conclude that the appropriate
transformation law is $e^{i\phi^{\prime a}
  X^a} = \Omega e^{i\phi^a X^a} $, but this is not so. Whilst we know
that $\Omega e^{i\phi^a X^a} $ is an element of $G$, we do not know
that we can write it in the form $e^{i\phi^{\prime a}
  X^a}$! In general it will take the form $e^{i(\phi^{\prime a}
  X^a + \psi^{\tilde{a}} T^{\tilde{a}}) }$. But there is an easy fix. Since $h\Phi_0 = \Phi_0$ for any $h
\in H$, we also have that $e^{i\phi^a X^a}\Phi_0 \mapsto e^{i\phi^{\prime a}
  X^a} \Phi_0\equiv \Omega e^{i\phi^a X^a} h \Phi_0$ and by choosing a
suitable $h$, we can remove the piece $e^{i \psi^{\tilde{a}} T^{\tilde{a}}
}$. Note that the required $h$ will depend on both $U$ and $\Omega$.

To summarise, the action of an element $\Omega$ of the symmetry group
$G$ on the fields $U(x)$ is given by
\begin{gather} \label{eq:nonl}
U(x) \mapsto \Omega U(x) h (\Omega, U(x)). 
\end{gather}
Note that this is a non-linear transformation (because of the
dependence of $h$ on $U$), which is why we say that the symmetry
$G$ is non-linearly realized on the fields $\phi(x)$.
\subsection{Building the EFT lagrangian}
We now want to build the most general action for the EFT, consistent
with the $G$ symmetry. This looks like a formidable task, because of
the complicated, non-linear way (\ref{eq:nonl}) in which the fields
$U(x)$ transform. But again there is a trick, which is to first build
objects that transform only under the subgroup $H$. To do so, consider
the object $U^{-1} \partial_\mu U$. Evidently, since $\Omega$ is
constant, this transforms as
\begin{gather} 
U^{-1} \partial_\mu U \mapsto h^{-1} (U^{-1}\partial_\mu U) h +
h^{-1}\partial_\mu h.
\end{gather}
Notice that the dependence on $\Omega$ has disappeared. Now,
$U^{-1} \partial_\mu U$ and the thing into which it transforms take
values in the Lie algebra of $G$. Thus we can decompose them in our
basis $\{T^{\tilde{a}}, X^a\}$ for the Lie algebra. We then have
that\footnote{Note that, even though $U=e^{i\phi^a X^a}$, it does not
  follow that $(U^{-1} \partial_\mu U)_H = 0$, because the generators
  $\{X^a\}$ do not close into themselves under the Lie bracket operation.}
\begin{gather} 
U^{-1} \partial_\mu U = (U^{-1} \partial_\mu U)_H + (U^{-1} \partial_\mu U)_X
\end{gather}
and we can decompose the transformation law as
\begin{align} \label{eq:split}
(U^{-1} \partial_\mu U)_X &\mapsto h^{-1} (U^{-1}\partial_\mu U)_X h , \\
(U^{-1} \partial_\mu U)_H &\mapsto h^{-1} (U^{-1}\partial_\mu U)_Hh +
h^{-1}\partial_\mu h.
\end{align}
These two pieces are more transparent: $(U^{-1} \partial_\mu U)_X$ is an object that transforms
homegeneously under $H$, while $(U^{-1} \partial_\mu U)_H$ transforms
like a covariant derivative under $H$.

We can now start to build invariants out of the coset fields using
$(U^{-1} \partial_\mu U)_X$ in the following way. We first note that
the fields $\phi^a$ actually transform as a representation under
$H$. What representation? Well, the elements of the Lie algebra of $G$
carry the adjoint representation of $G$ in general. But the fields
$\phi^a$ only transform under the subgroup $H \subset G$, so we should
first decompose the adjoint representation of $G$ into its irreducible
representations (irreps) under $H$. Finally, the fields $\phi^a$ are
projections on the subspace of the Lie algebra that is orthogonal to
the Lie algebra of $H$ and so we should remove the irrep that
corresponds to the adjoint irrep of $H$. We are left with the rep, $R$,
under which $(U^{-1} \partial_\mu U)_X$ transforms. Whenever the
tensor product of $n$ copies of $R$ contains a singlet, we can write
an invariant in the action involving $n$ copies of
$(U^{-1} \partial_\mu U)_X$.

In particular, it is a theorem that by taking just two copies of $R$,
we can form a singlet for every real irrep that is contained in $R$. 

For the MCHM, the adjoint rep of $SO(5)$ is 10 dimensional (the same as
the dimension of the Lie algebra). Under the $SO(4) \simeq SU(2) \times
SU(2)$ subgroup it
decomposes as $\bf{10} \rightarrow (\bf{3},\bf{1}) \oplus
(\bf{1},\bf{3}) \oplus (\bf{2},\bf{2})$. But $(\bf{3},\bf{1}) \oplus
(\bf{1},\bf{3})$ is just the adjoint rep of $SO(4) \simeq SU(2) \times
SU(2)$, so we see that the 4 fields $\phi^a (x)$ must transform as a
$(\bf{2},\bf{2})$ of $SU(2) \times
SU(2)$. This is precisely how the Higgs field of the SM transforms
under the custodial $SU(2)_L \times
SU(2)_R$ symmetry, and so the Goldstone bosons of the
$SO(5)/SO(4)$ non-linear sigma model have just the representation to
play the role of the SM Higgs!

So there is just one irrep in this case, and we can form just one
singlet that is quadratic in derivatives. It takes the form
\begin{gather}
-f^2 \mathrm{tr} (U^{-1} \partial_\mu U)_X^2 =
\frac{1}{2} \partial_\mu h^a \partial^\mu h^a + \dots
\end{gather}
where we have now written $U=e^{i h^a X^a/f}$, including a dimensionful scale $f$ so that the Higgs
field has the canonical unit dimension of a scalar field in 4-d. 
At leading order, we get precisely the kinetic terms of the Higgs
field in the SM. But at higher order we get terms with two derivatives
and higher powers of Higgs fields. These are, of course,
non-renormalizable, but we don't care any more, because we are doing
EFT. 

Note that we can also put in terms with more derivatives, by taking
more copies of $(U^{-1} \partial_\mu U)_X$. Each factor adds one more
derivative and in the EFT spirit that we expect all operators to
become equally important at the cut-off, they should be accompanied by
a factor of the cut-off $\La$. We thus get that
\begin{gather}\label{eq:exp}
\mathcal{L} \sim f^2 (U^{-1} \partial_\mu U)^2_X + \frac{f^2}{\La^2}
(U^{-1} \partial_\mu U)^4_X + \dots \, .
\end{gather}
\subsection{Estimate of the cut-off scale}
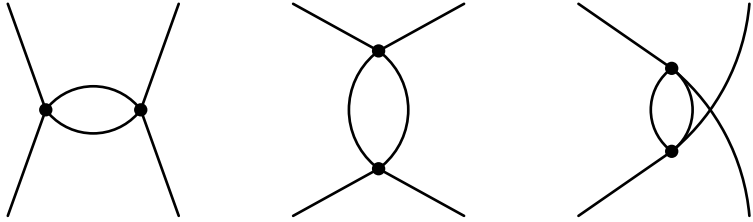
\begin{figure}
\centering
\begin{tabular}{c c c}

\begin{fmffile}{pi1}
\begin{fmfgraph}(80,80)
\fmfleft{i1,i2}
\fmfright{o1,o2}
\fmf{plain}{i1,v1}
\fmf{plain}{i2,v1}
\fmf{plain,left=0.5,tension=0.4}{v1,v2}
\fmf{plain,right=0.5,tension=0.4}{v1,v2}
\fmf{plain}{v2,o1}
\fmf{plain}{v2,o2}
\fmfdot{v1,v2}
\end{fmfgraph}
\end{fmffile}
&
\begin{fmffile}{pi2}
\begin{fmfgraph}(80,80)
\fmfleft{i1,i2}
\fmfright{o1,o2}
\fmf{plain}{i1,v1}
\fmf{plain}{i2,v2}
\fmf{plain,left=0.5,tension=0.4}{v1,v2}
\fmf{plain,right=0.5,tension=0.4}{v1,v2}
\fmf{plain}{v1,o1}
\fmf{plain}{v2,o2}
\fmfdot{v1,v2}
\end{fmfgraph}
\end{fmffile}
&
\begin{fmffile}{pi3}
\begin{fmfgraph}(80,80)
\fmfleft{i1,i2}
\fmfright{o1,o2}
\fmf{plain}{i1,v1}
\fmf{plain}{i2,v2}
\fmf{plain,left=0.5,tension=0.6}{v1,v2}
\fmf{plain,right=0.5,tension=0.6}{v1,v2}
\fmf{phantom}{v1,o1}
\fmf{phantom}{v2,o2}
\fmf{plain,right=0.2,tension=0.2}{v1,o2}
\fmf{plain,left=0.2,tension=0.2}{v2,o1}
\fmfdot{v1,v2}
\end{fmfgraph}
\end{fmffile}
\end{tabular}

\caption{One-loop contributions to $hh \rightarrow hh$.\label{fig:pipi}}
\end{figure}

Now, $f$ and $\Lambda$ are both dimensionful scales in the theory. We
have already seen that non-renormalizable terms involving extra powers
of the scalar fields are suppressed by powers of $f$, and so it must
be that $f$ is related to the cut-off $\La$, somehow. We shall now
argue that it is unreasonable to suppose that $\La$ is much greater
than $4\pi f$. The argument goes as follows. The leading order term in (\ref{eq:exp}) contains a
quartic interaction that goes like (Fourier transforming to momentum space)
\begin{gather} \label{eq:lo}
\frac{p^2 h^4}{f^2}.
\end{gather}
Consider the 3 one-loop diagrams in Figure~\ref{fig:pipi}, contributing to $hh \rightarrow hh$,
with two insertions of this vertex. By dimensional analysis, the loop integral na\"{\i}vely goes
like
\begin{gather}
\int d^4 k \frac{k^2 k^2}{f^4 k^2 k^2},
\end{gather}
which is
quartically
divergent. However, the group theory factors must be such that this
contribution gives zero when summed over the 3 diagrams, because
such a divergence would have to be cancelled by a counterterm of the
form $h^4$ with no derivatives, but this is not allowed by the
symmetry. When one works it out carefully, one finds that the
contribution is indeed zero by the Jacobi identity.
There is also a sub-leading piece which contains two powers of the
external momenta $p$ and goes like
\begin{gather}
p^2 \int d^4 k \frac{ k^2}{f^4 k^2 k^2}.
\end{gather}
This is quadratically divergent, but the divergence can be absorbed by
the term (\ref{eq:lo}) itself. 

Finally, there is a logarithmically divergent piece of size
\begin{gather}
\frac{p^4}{f^4 (4\pi)^2} \log \mu,
\end{gather}
where the $4\pi$ comes from the integration over a hypersphere. We get
similar contributions at tree-level from a piece 
\begin{gather} \label{eq:lo}
\frac{p^4 h^4}{\La^2 f^2}.
\end{gather}
coming from the second term in  (\ref{eq:exp}). Now, it cannot be the
case that $\Lambda \gg 4\pi f$, because if this were true for one
choice of renormalization scale, if would not be true for another that
differed by $O(1)$. Thus we conclude that $\Lambda \lesssim 4\pi f$.
\subsection{Pseudo-Goldstone bosons}
So far, we have built a consistent EFT of Goldstone bosons, in which
the usual problematic mass term operators are forbidden by
non-linearly realized symmetries. We have also found a specific model
in which the Goldstone bosons transform as a $(\bf{2},\bf{2})$ of an $SU(2) \times
SU(2)$ symmetry, just like the Higgs field of the SM. 

But we are still rather a long way from a model that can describe
Nature. Indeed, although we currently know rather little about the
Higgs boson, we do know that it is rather a long way from being a
Goldstone boson! It has a mass of 125 GeV, and it couples
to gauge fields and to SM fermions. Our Goldstone bosons have none of
these features, being massless, and coupled only to themselves, via
derivative interactions. 

To see how to solve these problems, we start by noting that there is
no way that $SO(4) \simeq SU(2) \times
SU(2)$, let alone $SO(5)$, can be an exact symmetry of Nature. We
already know, for example, that the custodial $SU(2) \times
SU(2)$ is only approximate, being broken both by Yukawa interactions
and by the gauging of the hypercharge. But if $SU(2) \times
SU(2)$ (and $SO(5)$) are only approximate, then the Goldstone bosons
of the $SO(5)/SO(4)$ model will only approximately be Goldstone bosons
and will only approximately be massless, etc. They will, to use the
lingo, become {\em pseudo-Goldstone bosons}.

So the question is: can we somehow break $SO(4)$ and $SO(5)$ in a
small way, by introducing gauge interactions and couplings of the
Goldstone bosons to fermions, and thus end up with something much
closer to the SM? 

The answer is: Yes, we can! I am going to show you how to do properly for the
gauge interactions only, and sketch how it goes for the couplings to
fermions. This sounds like a bit of a cop out, but I really am going
to do it properly for the gauge couplings, and indeed we will obtain a
result which cannot be found elsewhere in the literature.

So, let us attempt the following. Starting with the $SO(5)/SO(4)$
non-linear sigma model, we will try to gauge the $SU(2)_L \times
U(1)_Y$ subgroup of $SO(4) \simeq SU(2)_L \times
SU(2)_R$, where $Y = T_R^3$. We expect that, as a result, the
Goldstone bosons
will acquire a
potential (like the SM Higgs) and we shall derive its general form. 

We will do this by using the trick of spurions. Specifically, suppose
we wish to gauge a subgroup $K$ of the group $G$. (In the MCHM, $K = SU(2)_L \times
U(1)_Y$.) We will start by pretending that $K$ is not a subgroup of
$G$, but rather is separate, so the full theory has $K \times G$
invariance, where $K$ is a local symmetry (meaning that we have a
gauge field for it) and $G$ is the global symmetry of the sigma
model. We will then introduce a spurionic field $g^{A\alpha}$ (which we
call the {\em gauge coupling spurion}), which transforms as an adjoint
under $K$ (with index $\alpha$) and as an adjoint under $G$ (with index
$A$). We will declare that in the vacuum, $g^{A\alpha}$ has expectation
value given by\footnote{In fact, we can choose a different constant
  of proportion, $g$, for each simple factor in $K$, but we ignore
  this subtlety for now. We shall need it later, however, because $SU(2)_L \times
U(1)_Y$ has two simple factors.}
\begin{gather}
\label{eq:vev}
\langle g^{A\alpha} \rangle = g \delta^{A\alpha}.
\end{gather}

Now we can see how to write down a potential for the Goldstone
bosons. 
Under $G$, the field $g^{A\alpha}$ transforms as an adjoint. This is
conveniently expressed by defining $g^{\alpha} = g^{A\alpha} T^A$, s.\
t.\ 
the transformation law is $g^{\alpha} \mapsto \Omega g^\alpha
\Omega^{-1}$. We now observe that the object $\tilde{g}^\alpha \equiv
U^{-1} g^\alpha U \mapsto h^{-1} \tilde{g}^\alpha h$ and transforms
not under $G$, but under $H$, so we can easily build invariants from
it!

Now, $\tilde{g}^\alpha$ is an adjoint of $G$, so to see how it
decomposes under $H$, we just need to do the decomposition of the
adjoint of $G$ under $H$. For the MHCM, we get $\bf{10} \rightarrow (\bf{3},\bf{1}) \oplus
(\bf{1},\bf{3}) \oplus (\bf{2},\bf{2})$.

The object $\tilde{g}^\alpha$ still transforms under $K$, but we can
get a $K$ invariant by forming the quadratic object
$\tilde{g}^\alpha \tilde{g}^\alpha$. This transforms as the product of
two adjoints of $G$ and we already know that we can get one
$H$-invariant for each real irrep of $H$ that appears in the
decomposition of the adjoint
of $G$. However, the sum of all these terms is just the trace of
$\tilde{g}^\alpha \tilde{g}^\alpha$, which is a constant, independent
of the PGBs. Thus, we obtain our final result, which is that {\em the
  number of independent potential terms is one fewer than the number
  of real irreps of $H$ in the adjoint rep of $G$.} For the MCHM,
there are 3 real irreps, {\em viz.} $(\bf{3},\bf{1}),
(\bf{1},\bf{3}) $, and $ (\bf{2},\bf{2})$ and hence 2 independent
terms in the potential. I compute them in the Appendix. They are
\begin{gather}\label{eq:pots}
V(h) = 2A (3g^2 \cos^4 \frac{h}{2f} +  g^{\prime 2} \sin^4 \frac{h}{2f} )
+ 2B  (3g^2 \sin^4 \frac{h}{2f} +  g^{\prime 2} \cos^4 \frac{h}{2f} ),
\end{gather}
where, as always, $A$ and $B$ are arbitrary parameters in the EFT, to
be fixed by measurements.

A few remarks now follow.
 Firstly, note that these potential terms depend quadratically
on the gauge couplings. They are thus the dominant contributions for small couplings,
corresponding to a weak breaking of the $G$ symmetry. 

Secondly, there is a variant of the MCHM in which the $SO(4)$ symmetry
is enlarged to $O(4)$, so as to protect the theory from overly large
contributions to the decay rate for $Z \rightarrow b\overline{b}$
\cite{Agashe:2006at}. The reducible rep $(\bf{3},\bf{1}) \oplus
(\bf{1},\bf{3})$ of $SO(4)$ is actually an irrep of $O(4)$ \cite{Gripaios:2014pqa} and
so in this case there is just a single potential term, given by 
\begin{gather} \label{eq:inv1}
V(h) = A (3g^2 + g^{\prime 2}) \sin^2 \frac{h}{f}.
\end{gather}
This is the expression that you will find everywhere in the literature, even for
the $SO(4)$ case.
\subsection{Composite Higgs}
We are still quite a long way from a realistic composite Higgs model. 
For example, with $A,B >0$ in (\ref{eq:pots}), we have a minimum at
the origin, and so we can't break the electroweak symmetry as needed.

This can be fixed though, once we add another source of breaking by
coupling the pseudo-Goldstone bosons to fermions. 
These coupings must be present, because we know that the Higgs (which
is here part of the strongly coupled sector) couples to fermions (and
gives them mass after EWSB).
There are two ways in which we can imagine the couplings arising. The
first is much like the SM Yukawa couplings, in that the strong sector 
couples to fermion bi-linears. Schematically, 
\begin{gather} \label{old}
\mathcal{L} \supset \frac{q \mathcal{O}_h u^c
}{\Lambda^{d-1}} + \dots,
\end{gather}
 where $\mathcal{O}_h$ is some operator in the
strong sector of arbitrary dimension $d$ with the right quantum numbers to couple to SM
fermions. 

However, to this EFT lagrangian we should also add other operators that are compatible with the symmetries of the theory. Amongst these are
\begin{gather} \label{danger}
\mathcal{L} \supset \frac{q q q q}{\Lambda^{2}} + \Lambda^{4 - d'}\mathcal{O}_{h}^\dagger \mathcal{O}_{h}.
\end{gather}
The first of these is responsible for flavour changing neutral
currents; for these to be small enough, $\Lambda > 10^{3-5}$ TeV. But
then, in order to get a mass as large as that of the top from the
operator in (\ref{old}), we need to choose $d$ to be rather small: $d
\lesssim 1.2 - 1.3$ \cite{Luty:2004ye}. Next, we need to worry about
the second operator in (\ref{danger}). In order not to de-stabilize
the hierarchy, its dimension, $d'$, had better be greater than four,
rendering it irrelevant.\footnote{It is, perhaps, instructive to see
  how the hierarchy problem of the SM is cast in this language. There,
  $\mathcal{O}_{h}$ corresponds to the Higgs field $h$, with dimension
  close to unity, whilst $\mathcal{O}_{h}^\dagger \mathcal{O}_{h}$ is
  the Higgs mass operator, with dimension close to 2.} So what is the
problem? The limit in which $d \rightarrow 1$ corresponds to a free
theory (for which the operator $\mathcal{O}_h$ is just the Higgs field
$h$), and in that limit $d' \rightarrow 2d \rightarrow 2$. So in order to have an
acceptable theory, we need a theory containing a scalar operator
$\mathcal{O}_h$ (with the right charges) with a dimension that is
close to the free limit, but such that the theory is nevertheless
genuinely strongly-coupled, with the dimension of
$\mathcal{O}^\dagger_h \mathcal{O}_h$ greater than four. We have very
good evidence that such a theory cannot exist \cite{Rattazzi:2010yc}.

In the other approach, we imagine that the elementary fermions couple
linearly to fermionic operators of the strong sector \cite{Kaplan:1991dc}. Schematically, the lagrangian is
\begin{gather} \label{mix}
\mathcal{L} \sim   q \mathcal{O}_{q^c} +    u^c \mathcal{O}_{u} +   \mathcal{O}_{q^c} \mathcal{O}_{q} + \mathcal{O}_{u^c} \mathcal{O}_{u} +  \mathcal{O}_{q^c} \mathcal{O}_H \mathcal{O}_{u}
\end{gather}
(where I have left out the $\Lambda$s)
and the light fermion masses arise by mixing with heavy fermionic
resonances of the strong sector, which feel the electroweak symmetry
breaking. The beauty of this mechanism is that fermion masses can now
be generated by relevant operators ({\em cf.} the operator that
generates masses in (\ref{old}), which is at best marginal, since
$d>1$); this means that one can, in principle, send $\Lambda$ to
infinity and the problems with flavour physics can be
completely decoupled. There is even a further bonus, in that the light
fermions of the first and second generations, which are the ones that
flavour physics experiments have most stringently probed, are the ones
that are least mixed with the strong sector and the flavour-changing
physics that lies therein. In this model, the observed SM fermions are
mixtures of elementary and composite fermions, with the lightest
fermions being mostly elementary, and the top quark mostly
composite. The scenario therefore goes by the name of {\em partial
  compositeness}.

It turns out (see, {\em e.\ g.}, \cite{Contino:2010rs}) that the fermions can give negative corrections
to the mass-squared in the Higgs potential, and thus result in
EWSB. Since the top quark Yukawa is somewhat bigger than the gauge
couplings, this is (at least na\"{\i}vely) the most likely outcome.

We now have something approaching a realistic model of EWSB via strong
dynamics. Having built it up, we should now do our best to knock it
down. 

A first problem is that no one actually knows how to get a pattern of
$SO(5) \rightarrow SO(4)$ global symmetry breaking out of an explicit
strongly-coupled gauge theory coupled to fermions.\footnote{The
  breaking $SO(6) \rightarrow SO(5)$ \cite{Gripaios:2009pe} is
easier to achieve, since $SO(6) \simeq SU(4)$, and unitary
groups are easier to obtain.} 

A second problem is the $S$-parameter. One can argue (see \cite{Gripaios:2015gxa} for
  more details) that the
necessary suppression can be obtained if $v$ turns out to be somewhat
smaller than $f$, the scale of strong dynamics. Well, $v$ is obtained
by minimizing the Higgs potential $V(h)$, which contains contributions
of very roughly equal size, but opposite in sign, from the top quark
and gauge bosons. Thus it is possible to imagine that there is a
slight cancellation due to an accident of the particular strong
dynamics, such that the $v$ that emerges is small enough. A measure of
the required tuning is $\frac{v^2}{f^2}$, and the observed $S$-parameter requires
tuning at the level of ten {\em per cent} or so.

The third problem concerns flavour physics. To argue, as we have done above,
that the flavour problem can be decoupled, is not the same as arguing
that it {\em is} solved. To do that, one
needs to find an explicit model which possesses all the required
operators, with the right dimensions. Needless to say, our ignorance
of strongly-coupled dynamics means we have no idea whether such a
model exists. Certainly, in all cases that have been studied (either
models with large rank of the gauge group, or lattice studies), there
{\em is} a problem with flavour constraints. 

Despite these problems, composite Higgs models seem just as good (or
just as bad) as solutions to the hierarchy problem as supersymmetric
models, and so they deserve thorough investigation at the LHC. This
itself is not so easy to do. Na\"{\i}vely, the obvious place to look
for deviations is in the Higgs sector itself, for example in the
couplings of the Higgs boson to other particles. However, we know that
(since such models reproduce the SM in the limit $v^2/f^2 \rightarrow
0$) the deviations must be proportional to $v^2/f^2$ and hence at
most 10 \% or so. Such deviations are hard to see at the LHC,
and even at a future $e^+ e^-$ collider. Perhaps a better way is to
look for the composite partners of the top quark, which must be not
too heavy in order to reproduce the observed Higgs mass. Many
suggestions for how to do so have been put forward and the experiments
are beginning to implement them. See, {\em e.g.} \cite{DeSimone:2012fs,Gripaios:2014pqa} and
refs. therein for more details.
\section{Third example: The quantum theory of fluids}
In this lecture, I describe a rather different EFT, namely that of a
perfect fluid \cite{Gripaios:2014yha}.\footnote{We take the fluid to be perfect because otherwise we expect to see
dissipative or viscous behaviour. But the quantum theory would then
presumably be non-unitary.} This is of interest in its own
right, since classical fluid phenomena are among the most rich and
fascinating in Nature. We will show that the quantum EFT based on the
same degrees of freedom and symmetries is a sensible
theory. Presumably, the quantum phenomena of this theory are even more
fascinating than those of a classical fluid, and so it is of interest
to explore the predicitions of the theory and search for evidence of
systems that behave in this way in Nature.

The fluid EFT is also of interest because it is a rather non-trivial
example of an EFT. In one sense, it is just a theory of Goldstone
bosons like the non-linear sigma models we discussed in the last
lecture. But it is more complicated, because the symmetry group is
infinite-dimensional and because the non-linearly realized symmetries
include spacetime symmetries. In particular, Lorentz invariance is
non-linearly realized in the ground state and so we must take care in
formulating the EFT. 
\subsection{Parameterization of a perfect fluid}
We begin by discussing how to parameterize a fluid and its
dynamics. Let the fluid occupy some spatial manifold $M$
(e.g. $\mathbb{R}^2$) and choose some co-ordinates $x^i$ thereon. At $t=0$,
we can label each fluid particle by the co-ordinates of the point in
$M$ that it occupies. Call these Eulerian co-ordinates, $\phi^i$. As
time evolves, the fluid particle will move around in $M$ and we can
denote its position at time $t$ by $x^i (\phi^j,t)$. Alternatively
(assuming the map is invertible, which requires that the fluid does
not cavitate or interpenetrate), then we can also describe the fluid's
configuration by the map $\phi^i (x^j,t)$. We choose to think about
things this way, since we can then think of the $\phi^i$ as 2 scalar
fields living in spacetime $(x^j,t)$.  

Note that the classical ground state corresponds to each fluid
particle sitting at rest. So the classical ground state is given by
$\phi^i = x^i$. Later, it will be useful to consider small
fluctuations about the classical ground state, which we write as 
$\phi^i = x^i + \pi^i (x^j, t)$. Again, the $\pi^i$ can be thought of
as 2 scalar fields on spacetime.
\subsection{Action principle and classical fluid dynamics}
We have now identified the degrees of freedom for the EFT. We next
wish to identify the symmetries. We do this essentially by guessing
and showing that the resulting action reproduces the behaviour of a perfect
fluid in the classical limit. 

The action of a fluid has been known for a long time \cite{herglotz},
but it is hard to find in the fluid mechanics textbooks,\footnote{One
  place you can find it is in \cite{soper1975}.} where it is usual to
derive the fluid equations of motion from conservation of energy and
momentum instead.

The action is the most general one consistent with the following
symmetries. Firstly, we require that the action be invariant under
Poincar\'{e} transformations of $x$. This is because we expect the
underlying dynamics of a fluid to be Poincar\'{e} invariant (though
its ground state, with the fluid sat still in some frame, is
not!). Secondly, we require that the action be invariant under 
area-preserving diffeomorphisms of the co-ordinates $\phi$. This is
because such transformations simply correspond to different labellings
of the fluid particles or `remixings' of the fluid.

At leading order, the lagrangian (in 2+1-d spacetime) can then be
written as\footnote{Our metric is now mostly-plus. If you don't like
  it, sue me!}
\begin{gather}
\mathcal{L} = - w_0 f(\sqrt{B}),
\end{gather}
where $B = \mathrm{det} \ \pt_\mu \phi^i \pt^\mu \phi^j$, $f$ is
any function s.\ t.\ $f^\prime (1)=1$, and $w_0$ sets the overall
dimension. It is easy to check that $B$ is indeed invariant under the
desired symmetries, and therefore so is $f(\sqrt{B})$. 

Since the theory is invariant under spacetime translations, Noether's
theorem tells us that there is a conserved energy-momentum tensor.
It may be written as 
\begin{gather} \label{eq:fl}
T_{\mu \nu} = (\rho + p) u_\mu u_\nu + p \eta_{\mu \nu},
\end{gather}
where
\begin{align}
\rho &= w_0 f ,\\ 
p &=  w_0 ( \sqrt{B} f^\prime -f), \\ 
u^\mu &= \frac{1}{2\sqrt{B}}\epsilon^{\mu \alpha \beta} \epsilon_{ij}
\pt_\alpha \phi^i \pt_\beta \phi^j .
\end{align}
Eq. (\ref{eq:fl}) is, of course, the standard form for the
energy-momentum tensor of a fluid with density $\rho$, pressure $p$
and 3-velocity $u^\mu$ of the fluid particles. Thus our action does indeed
reproduce a classical fluid. Note that different choices for the
function $f$ lead to different relations between the pressure and
density of the fluid, meaning that we are able to describe a fluid
with an arbitrary equation of state. Note also that these physical
quantities are invariant under the area-preserving diffeomorphisms of
$\phi$, which correspond to physically-equivalent relabellings of the
fluid particles. 
\subsection{Effective field theory: IR divergences}
Let's now try to study small fluctuations about the classical ground
state and see if we can make a consistent EFT. So, expanding
$\phi^i = x^i + \pi^i$, we get
\begin{multline}
\label{eq:2dlag}
\mathcal{L} = \frac{1}{2} (\dot\pi^2 - c^2 [\pt\pi]^2) - \frac{(3c^2 + f_3)}{6} [\pt\pi]^3 + \frac{c^2}{2}  [\pt\pi] [\pt\pi^2] + \frac{(c^2 +1)}{2}  [\pt\pi] \dot\pi^2 - \dot\pi \cdot \pt \pi \cdot \dot\pi \\
-\frac{(f_4 + 3c^2 + 6f_3)}{24} [\pt\pi]^4 + \frac{(c^2 + f_3)}{4} [\pt\pi]^2[\pt\pi^2] - \frac{c^2}{8} [\pt\pi^2]^2 
+ \frac{(1-c^2)}{8} \dot\pi^4 -c^2 [\pt\pi] \dot\pi \cdot \pt\pi \cdot \dot\pi \\- \frac{(1-3c^2-f_3)}{4} [\pt\pi]^2 \dot\pi^2 
+ \frac{(1-c^2)}{4} [\pt\pi^2] \dot\pi^2 + \frac{1}{2} \dot\pi \cdot
\pt\pi \cdot \pt\pi^T \cdot \dot\pi + \dots,
\end{multline}
where $f_n \equiv d^n f/d\sqrt{B}^n |_{B=1}$, $c \equiv \sqrt{f_2}$, and $[\pt
\pi]$ is the trace of the matrix $\pt^i \pi^j$, {\em \&c}.

This is the sort of expression that is liable to give one a heart
attack, so let's break it down and approach it bit by bit. 

The first thing to notice is that all the terms have derivatives in
them, either with respect to space or time or both. So this is a
theory of Goldstone bosons, albeit a funny one.

Next, let's look at the quadratic piece in $\pi$. It is just
\begin{gather}
\frac{1}{2} (\dot\pi^2 - c^2 [\pt\pi]^2).
\end{gather}
What do we learn from this? Recall that there are two scalar degrees
of freedom, $\pi^i$, with $i \in \{1,2\}$. Suppose we choose a
fluctuation mode with energy $\om$ and wavevector $(k^1,k^2) =
(k,0)$. For the mode $\pi^1$, which is longitudinally polarised (it's
in the same direction as $\vec{k}$), the lagrangian is just
 \begin{gather}
\frac{1}{2} (\om^2 - c^2 k^2 )(\pi^1)^2,
\end{gather}
meaning that the dispersion relation for longitudinally polarised
modes in the fluid is just $\om^2 (k) = c^2k^2$. Longitudinally polarised
excitations of a fluid are called sound waves, and we learn that they
have speed $c$. This is why we defined $c \equiv \sqrt{f_2}$ above. 

For the mode $\pi^2$, which is transversely polarised (it's
orthogonal to $\vec{k}$), we instead get
\begin{gather}
\frac{1}{2} \om^2 (\pi^2)^2,
\end{gather}
meaning that the dispersion relation is just $\om^2 (k) = 0$. This
looks a bit odd at first, but (at least classically) there is no
problem. A transversely polarized small fluctuation is just an
infinitesimal version of a fluid vortex. The dispersion relation says
firstly that the energy of such vortices is independent of $k$ -- that is,
independent of the size of the vortex. It also says that the energy of
such a vortex is zero. Both of these statements make sense. Indeed (as
you can check for yourself whilst sitting in the bath) it is possible
to firstly make a vortex of arbitrary size with arbitrarily low
energy, simple by placing ones hands the required distance apart and
stirring the bathwater arbitrarily slowly. 

So, classically, there is no problem. But there is a problem when we
start trying to do EFT. In particular, the spacetime propagator for
the transverse modes is given by $\int
d \om d^2k \; e^{i (\om t + k\cdot x)} /\om^2$ and this is undefined, because of
the pole at $\om = 0$. Note how this differs from normal scalar field
theory, where the propagator is given by $\int
d \om d^2k \; e^{i (\om t + k\cdot x)} /(\om^2 + k^2)$, which is
perfectly well-defined.

This obstruction to quantization was noted some time ago by Endlich
{\em et al.} \cite{Endlich:2010hf}. They tried to fix it up by adding
a small sound speed $c_T$ for the transverse modes, computing $S$-matrix
elements, and then sending $c_T \rightarrow 0$. Unfortunately they
found that everything they computed diverged as $c_T \rightarrow 0$.

In fact, it is not hard to see why this is the case. The fields
$\pi^i$, just like the fields $\phi^i$, are not physical. They
correspond to arbitrary labellings of the fluid elements. We cannot,
therefore, reasonably insist that correlation functions of them make
sense. The correlation functions of them are, in fact, infra-red
divergent (the propagator, for example, diverges because of the pole
at $\om = 0$) and this is a common feature in theories that are
formulated in terms of unphysical degrees of freedom. The most obvious
example occurs in gauge theories, where the gauge fields themselves
are unphysical, and indeed we find IR divergences whenever we attempt
to calculate correlation functions of the gauge fields. Another
example arises in non-linear sigma models, exactly like those we
studied in the last lecture, but in 2-d, where the propagator is $\int
d \om d k \; e^{i (\om t + k\cdot x)} /(\om^2 + k^2)$ and is also IR
divergent.
In these other theories, the solution to the problem of IR divergences
is well known: they cancel when we compute correlation functions of
physical quantities, such as gauge invariants in the case of gauge
theories. 

Does this work for the fluid as well? Indeed it does. There, the
physical quantities are invariants under area-preserving
diffeomorphisms, like the fluid's density, pressure, and 3-velocity. 
The latter is given in terms of the $\pi$ fields, at leading order, by
$u^i \propto \dot{\pi}^i$ and $u^0 \propto [\pt \pi]$ and it is almost
trivial to check that the 2-point functions of these physical
quantities are well behaved \cite{Gripaios:2014yha}. 

Indeed, we find that 
\begin{align}
\label{eq:2pt}
\langle [\pt \pi][\pt \pi] \rangle &= 
\frac{ik^2}{\om^2 -c^2 k^2}, \nonumber \\
\langle \dot{\pi}^i [\pt \pi] \rangle &= \frac{i\om
  k^i}{\om^2 -c^2 k^2},
\nonumber \\
\langle \dot{\pi}^i \dot{\pi}^j \rangle &= 
i\delta^{ij} + \frac{ic^2 k^i k^j}{\om^2 -c^2 k^2}.
\end{align}
The only poles are at $\om = c k$ 
and the disappearance of poles at $\om = 0$ implies that the spacetime
Fourier transforms are well-defined. 

The calculations for higher-point, tree-level correlation functions
are much more involved, but the cancellations have been checked in a
number of cases. See \cite{Gripaios:2014yha} for more details. 
\subsection{Effective field theory: UV divergences}
Now we have got the IR divergences under control, we can look at the
UV behaviour of the EFT. We would like to show that the EFT expansion
makes sense, in that there is a regime of large distance and time
scales (not necessarily the same, since the ground state is not
Lorentz-invariant) in which the effects of higher dimension operators
and loops are suppressed. 

\begin{figure}
\begin{center}
\begin{fmffile}{loop_feynmp}
\begin{fmfgraph*}(80,80)
\fmfleft{l1}
\fmfright{r1}
\fmf{plain}{l1,v1}
\fmf{plain}{v2,r1}
\fmf{phantom}{v1,v2}
\fmffreeze
\fmf{plain,left=1}{v1,v2}
\fmf{plain,right=1}{v1,v2}
\end{fmfgraph*}
\end{fmffile}
\begin{fmffile}{loop2_feynmp}
\begin{fmfgraph*}(80,80)
\fmfleftn{l}{21}
\fmfrightn{r}{21}
\fmftop{t1}
\fmfbottom{b1}
\fmf{phantom,tension=0.1}{t1,vdummy1}
\fmf{phantom,tension=0.1}{b1,vdummy2}
\fmf{plain,tension=0.5}{l10,vdummy2}
\fmf{plain,left=0.2}{vdummy1,v1,vdummy2}
\fmf{plain,tension=0.5}{vdummy1,l12}
\fmf{plain,tension=0.8}{v1,r11}
\end{fmfgraph*}
\end{fmffile}

\begin{fmffile}{loop3_feynmp}
\begin{fmfgraph*}(80,80)
\fmfleftn{l}{21}
\fmfrightn{r}{21}
\fmftop{t1}
\fmfbottom{b1}
\fmf{phantom,tension=0.1}{t1,vdummy1}
\fmf{phantom,tension=0.1}{b1,vdummy2}
\fmf{plain,tension=0.5}{r10,vdummy2}
\fmf{plain,right=0.2}{vdummy1,v1,vdummy2}
\fmf{plain,tension=0.5}{vdummy1,r12}
\fmf{plain,tension=0.8}{v1,l11}
\end{fmfgraph*}
\end{fmffile}
\begin{fmffile}{loop4_feynmp}
\begin{fmfgraph*}(80,80)
\fmfleftn{l}{21}
\fmfrightn{r}{21}
\fmf{plain}{r10,l10}
\fmf{plain}{r12,l12}
\end{fmfgraph*}
\end{fmffile}
\end{center}
\caption{Diagrams for the correlator $\langle (\sqrt{B} u^0 - 1) (\sqrt{B} u^0 - 1) \rangle$.\label{fig:loopdiags}}
\end{figure}
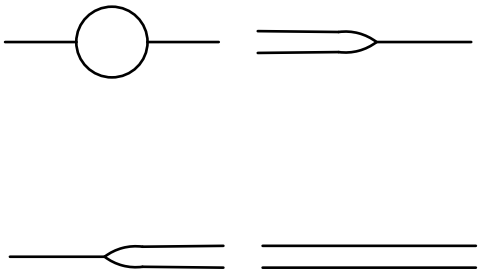

To check this, we compute the one-loop contribution to the 2-point
function of the observable $\sqrt{B} u^0 - 1\equiv [\pt \pi] + \frac{1}{2} ([\pt\pi]^2 - [\pt\pi^2])$. The diagrams, shown in Fig.~\ref{fig:loopdiags}, feature both IR and UV divergences, which
we regularize by computing the integrals in $D=1+2\epsilon$ time- and
$d=2+2\epsilon$ space-dimensions. We wish to show that the UV
divergences can be absorbed in higher order counterterms and that
the expansion in energy and momenta is valid in some non-vanishing
region.

Fortunately, in the case at hand, we can be sure that the answer must
be finite as $\epsilon \rightarrow
0$ (if the theory is consistent).
This is because we can show by dimensional analysis that there can be no counterterms! Indeed
the Feynman rules that follow from (\ref{eq:2dlag}) imply that the 1-loop diagrams
must contain 3 more powers of energy or momentum than the tree-level
diagrams (because every $\pi$ is always accompanied by a derivative). Now, since the correlator can only be a function of $K^2$
(where $ic K\equiv \om$)
and $k^2$ (by time-reversal and rotation invariance, respectively),
the 1-loop contribution necessarily contains radicals of $K^2$
and $k^2$. But higher order
counterterms can only yield tree-level contributions that are rational
functions of $K^2$
and $k^2$ and so cannot absorb divergences in the 1-loop
contribution. 

The actual computation is a pig to do, but the answer is pretty
simple. One gets \cite{Gripaios:2014yha}
\begin{multline*}
\frac{9Kk^6(1+c^4)}{64(K^2+k^2)^2} 
-\frac{k^4}{1024 c^4 (K^2 +k^2)^\frac{5}{2}}\\
\times \Big[ c^4(1-c^2)^2 (19k^4 - 4K^2 k^2 + K^4)\\
-2f_3c^2 (1+c^2) k^2 (5k^2 +14K^2)
+ f_3^2(3k^4 + 8K^2 k^2 +8 K^4)
\Big], \nonumber
\end{multline*} 
which is indeed finite, as consistency demands. Moreover, there are no
poles at $K=0$ and the Fourier transform is well defined. Note that
the only pole is at $K^2 + k^2 = 0 \implies \om^2 = c^2 k^2$, implying
that the sound speed is not renormalized at one-loop.

\begin{figure}
\includegraphics{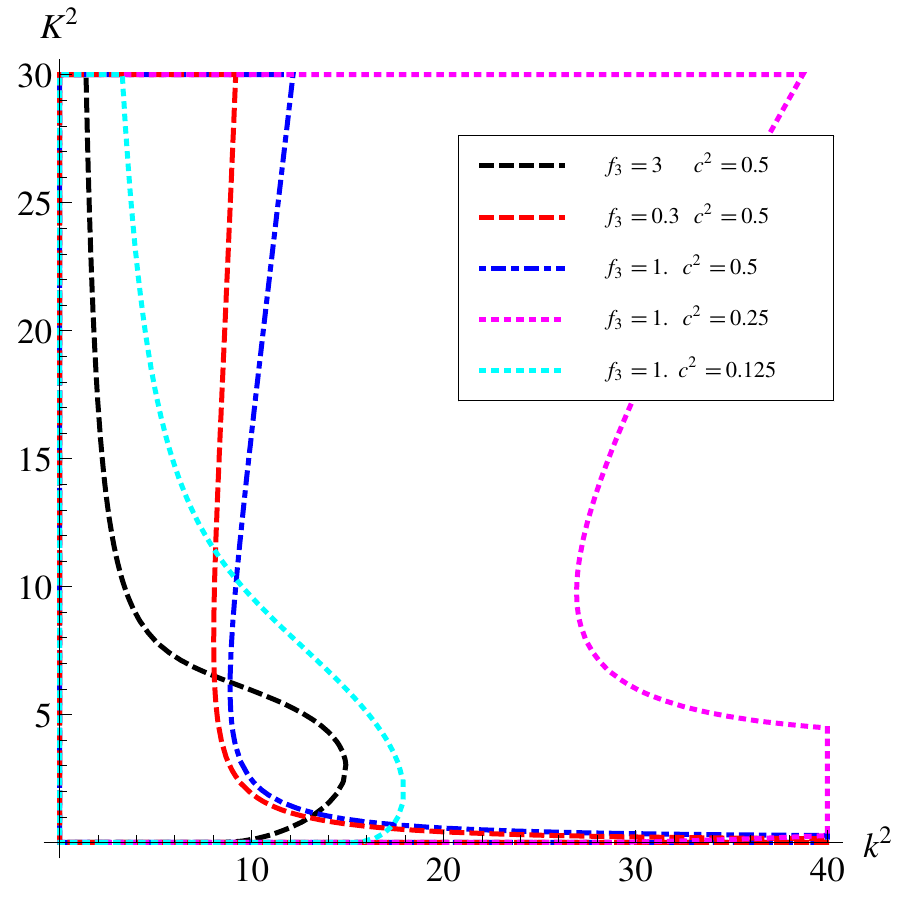}
\caption{Contours of equal 1-loop and tree-level contributions to
  $\langle (\sqrt{B} u^0 - 1)
  (\sqrt{B} u^0 - 1) \rangle$, for various values of $c$ and
  $f_3$. Figure reproduced from \cite{Gripaios:2014yha}.\label{fig:valid}}
\end{figure}

We can use this formula to estimate the region of validity
of the EFT
expansion in energy-momentum, by comparing the absolute values of the tree-level and
1-loop results. The estimate depends on the values of the
$O(1)$ coefficients $c^2$ and $f_3$; results for
typical values (in units of the overall scale $w_0$) are shown in
Fig.~\ref{fig:valid}. As required, the loop contribution is always
smaller than the tree-level one for large enough distances and times,
suggesting that the EFT expansion does indeed make sense. Notice also
that there is no suggestion that the locus of convergence of the EFT
expansion is some simple curve, such as $\La_\om = c \La_k$!
\appendix
\section{Appendix: Potential terms in the minimal composite Higgs model }
Here, we explicitly construct the potential invariants that arise from
gauge the electroweak subgroup in the $SO(5)/SO(4)$ MCHM. We use
the basis for $SO(5)$ generators given in \cite{Agashe:2004rs}. We denote the generators in $H=SU(2)_L \times SU(2)_R$ as
$L^a,R^b$ and the remaining generators as $X^c$.

The coset representative is, in an obvious notation,
\begin{gather}
U = e^{i \sqrt{2} h \cdot X} = \begin{pmatrix} 1 & 0 \\ 0 &
  1 \end{pmatrix} 
+ \frac{\sin h}{h} \begin{pmatrix} 0 & h \\ -h^T & 0 \end{pmatrix}
+ \frac{\cos h - 1}{h^2} \begin{pmatrix} hh^T & 0 \\ 0 & h^2 \end{pmatrix}.
\end{gather}
Now, the gauge coupling spurion is an adjoint of $G$ (and $K$), so we
may represent the $G$-action by $\Omega: g \rightarrow \Omega g
\Omega^{-1}$. The combination $\tilde{g} \equiv U^{-1} g U$, then transforms
(reducibly) under $H$ alone.

As described in the main text, for each irreducible representation under $SO(4)$, we
can build an $SO(5)$-invariant. To do so, we need to reduce
$\tilde{g}$ into its components carrying irreps. of $SO(4)$. This is easy: $\tilde{g}$
is an element of the Lie algebra of $SO(5)$, so we may expand it as
\begin{gather}
\tilde{g}^{\alpha} = \lambda^{A \alpha} L^A + \rho^{B \alpha}R^B + \mu^{C \alpha}X^C
\end{gather}
 and the three irreps. of $SO(4)$ are carried
precisely by the projections of  $\tilde{g}$ onto the
subalgebras corresponding to $SU(2)_L$
and $SU(2)_R$, together with their complement in $SO(5)$. The projection itself
is trivial, since our basis of generators for $SO(5)$ was chosen to be
orthogonal with respect to the trace operation. Thus,
\begin{align}
\lambda^{A \alpha} &= \mathrm{tr} L^A \tilde{g}^{\alpha}\\
\rho^{A \alpha} &= \mathrm{tr} R^A \tilde{g}^{\alpha}
\end{align}
each transform amongst themselves under $SO(4)$. (So, of course, does
the projection onto the $X$ subalgebra, but the sum of all three
invariants is coset-independent.)

Let us now consider, as a first example, gauging the whole of $SO(4)$, but with different
couplings, $g_L$ and $g_R$, for the two simple subgroups. We label the
generators of $K = SO(4)$ by $\{L^{\prime
  \alpha},R^{\prime \beta}\}$, such that the VEV of the gauge coupling
spurion may be written as
\begin{gather}
\langle g\rangle \equiv \langle g^\gamma T^{\prime \gamma} \rangle =
 g_L L^{\prime\alpha} \otimes L^\alpha + 
 g_R R^{\prime\beta} \otimes R^\beta,
\end{gather}
with
\begin{gather}
\langle g^\gamma \rangle =
 \begin{cases}
g_L L^\alpha, \;\mathrm{if}  \gamma \in \{\alpha\}  \\
g_R R^\beta, \;\mathrm{if}  \gamma  \in \{\beta\} \\
0, \; \mathrm{else}.
\end{cases}
\end{gather}

Thus, we find that
\begin{align}
\lambda^{A \alpha} &=  g_L\mathrm{tr} L^A U^{-1} L^{\prime\alpha} U + 
g_R \mathrm{tr} L^A U^{-1} R^{\prime\alpha} U \equiv \lambda_{L^\prime}^{A
  \alpha} +  \lambda_{R^\prime}^{A
  \alpha} \\
\rho^{A \alpha} &=  g_L\mathrm{tr} R^A U^{-1} L^{\prime\alpha} U +
g_R\mathrm{tr} R^A U^{-1} R^{\prime\alpha} U \equiv \rho_{L^\prime}^{A
  \alpha} +  \rho_{R^\prime}^{A
  \alpha} 
\end{align}

This is starting to look exceedingly unpleasant, but salvation comes
in the form of a {\em deus ex Mathematica}:
\begin{align}
\lambda_L^{A \alpha} \lambda_L^{A \beta} &=  g_L^2 \cos^4 \frac{h}{2}
\delta^{\alpha \beta}\\
\lambda_R^{A \alpha} \lambda_R^{A \beta} &=  g_R^2 \sin^4 \frac{h}{2} \delta^{\alpha \beta}\\
\rho_L^{A \alpha} \rho_L^{A \beta} &=  g_L^2\sin^4 \frac{h}{2} \delta^{\alpha \beta}\\
\rho_R^{A \alpha} \rho_R^{A \beta} &=  g_R^2\cos^4 \frac{h}{2} \delta^{\alpha \beta}
\end{align}

The two invariants that can appear in the Higgs potential are then given by
\begin{align}\label{eq:inv}
\langle \lambda^{A\alpha}  \lambda^{A \alpha} \rangle &= \mathrm{tr}
\lambda \lambda^T 
\propto 3g_L^2 \cos^4 \frac{h}{2} + 3g_R^2 \sin^4 \frac{h}{2}\\
\langle \rho^{A \alpha}  \rho^{A \alpha} \rangle &= \mathrm{tr} \rho \rho^T 
\propto 3g_R^2 \cos^4 \frac{h}{2} + 3g_L^2 \sin^4 \frac{h}{2}
\end{align}
 
A similar computation with 
\begin{gather}
\langle g\rangle \equiv \langle g^\gamma T^{\prime \gamma} \rangle =
 g L^{\prime\alpha} \otimes L^\alpha + 
 g^\prime R^{\prime 3} \otimes R^3,
\end{gather}
yields (\ref{eq:pots}).
\providecommand{\href}[2]{#2}\begingroup\raggedright\endgroup
\end{document}